\documentclass[11pt,prd,onecolumn,amsmath,amssymb,aps,floats,floatfix]{revtex4-2}
\usepackage[colorlinks=true,urlcolor=blue,anchorcolor=blue,citecolor=blue,filecolor=blue,linkcolor=blue,menucolor=blue,linktocpage=true]{hyperref} 

\usepackage[inline]{enumitem}
\usepackage[multidot]{grffile}  
\usepackage{dcolumn}
\usepackage{bm}
\usepackage{amsmath}
\usepackage{amsfonts}
\usepackage{amssymb}
\usepackage{color}
\usepackage[table]{xcolor}
\usepackage{float}
\usepackage{latexsym}
\usepackage{slashed} 
\usepackage{pstricks}
\usepackage{indentfirst}
\usepackage{mathrsfs}
\usepackage{multirow}
\usepackage{epsfig,psfrag}
\usepackage{graphicx}
\usepackage{enumitem}
\usepackage{geometry,amssymb,yfonts}
\usepackage{yhmath}
\usepackage{anysize}
\usepackage{subfigure}
\usepackage{mathtools}
\usepackage{setspace} 
\usepackage[utf8]{inputenc} 
\usepackage[scientific-notation=true]{siunitx} 

\graphicspath{{fig/}}

\allowdisplaybreaks 

\newcommand{\Uone}{\mathrm{U}(1)}
\newcommand{\UoneX}{\mathrm{U}(1)_\mathrm{X}}
\newcommand{\UoneY}{\mathrm{U}(1)_\mathrm{Y}}
\newcommand{\Uonebml}{\mathrm{U}(1)_{B-L}}
\newcommand{\SUtwoL}{\mathrm{SU}(2)_\mathrm{L}}

\begin{document}

\title{Ultraviolet completion of pseudo-Nambu-Goldstone dark matter with a hidden U(1) gauge symmetry}

\author{Dan-Yang Liu}
\author{Chengfeng Cai}
\author{Xue-Min Jiang}
\author{Zhao-Huan Yu}\email[Corresponding author. ]{yuzhaoh5@mail.sysu.edu.cn}
\author{Hong-Hao Zhang}\email[Corresponding author. ]{zhh98@mail.sysu.edu.cn}
\affiliation{School of Physics, Sun Yat-Sen University, Guangzhou 510275, China}

\begin{abstract}
We propose an ultraviolet completion model for pseudo-Nambu-Goldstone dark matter with a hidden $\mathrm{U}(1)$ gauge symmetry. Compared to previous studies, this setup is simpler, introducing less interactions. Dark matter scattering off nucleons is highly suppressed by the ultraviolet scale and direct detection constraints can be easily evaded. The kinetic mixing between the hidden $\mathrm{U}(1)$ and the $\mathrm{U}(1)_\mathrm{Y}$ gauge fields would lead to dark matter decays. We find that the current bound on the dark matter lifetime implies that the ultraviolet scale should be higher than $10^{10}~\mathrm{GeV}$. The phenomenological constraints from the 125 GeV Higgs measurements, the dark matter relic density, and indirect detection of dark matter annihilation are also investigated.
\end{abstract}

\maketitle
\tableofcontents

\clearpage

\section{Introduction}

The cosmological abundance of dark matter (DM) can be naturally explained by neutral weakly interacting massive particles (WIMPs) which are thermally produced in the plasma and subsequently freeze out as the Universe expands~\cite{Bertone:2004pz, Feng:2010gw, Young:2016ala}.
The crucial ingredient in this argument is the weak interaction strength of WIMP annihilation into standard model (SM) particles at the freeze-out epoch, implying that WIMP dark matter is quite promising to be probed in current direct detection experiments.
Although the direct detection sensitivity has been tremendously improved in the recent two decades, no robust signal has been found, suggesting severe constraints on the WIMP-nucleon scattering cross section~\cite{PandaX-4T:2021bab, Aalbers:2022fxq}.
This situation makes the WIMP paradigm questionable.

Nonetheless, an annihilation cross section of weak strength does not necessarily result in a WIMP-nucleon scattering cross section of the same strength.
It is possible to greatly suppress the scattering process in direct detection without affecting the annihilation processes at the freeze-out epoch.
An elegant way to realize it is to assume that the WIMP is a pseudo-Nambu-Goldstone boson (pNGB) whose interactions are momentum-suppressed~\cite{Gross:2017dan, Azevedo:2018exj, Ishiwata:2018sdi, Huitu:2018gbc, Alanne:2018zjm, Kannike:2019wsn, Karamitros:2019ewv, Cline:2019okt, Jiang:2019soj, Ruhdorfer:2019utl, Arina:2019tib, Abe:2020iph, Okada:2020zxo, Glaus:2020ihj, Abe:2020ldj, Zhang:2021alu, Abe:2021byq, Okada:2021qmi, Abe:2021jcz, Abe:2021vat, Biekotter:2021ovi, Zeng:2021moz, Cai:2021evx, Mohapatra:2021ozu, Darvishi:2022wnd, Abe:2022mlc, Biekotter:2022bxp}.
Since direct detection experiments basically operate at zero momentum transfer, the WIMP-nucleon scattering cross section totally vanishes at tree level~\cite{Gross:2017dan}, evading the direct detection constraints.

The original pNGB DM model~\cite{Gross:2017dan} introduces a complex scalar $S$, which is a SM gauge singlet.
The Lagrangian respects a $\Uone$ global symmetry $S \to e^{i\alpha} S$, except for a quadratic term $\mu'^2_S (S^2 + S^{\dag 2})/4$, which softly breaks the $\Uone$ symmetry into a $Z_2$ symmetry.
After the $\Uone$ spontaneous breaking, the imaginary part of $S$ becomes a stable pNGB, which has a mass $\mu'_S$ and acts as the WIMP with a vanishing tree-level WIMP-nucleon scattering amplitude.
Such a soft breaking term is \textit{ad hoc}.
Other soft breaking terms, such as a trilinear term $\propto S^3 + S^{\dag 3}$, would spoil the vanishing scattering amplitude.
Therefore, it demands an appropriate ultraviolet (UV) completion to realize only this quadratic soft breaking term~\cite{Gross:2017dan}.

A possible UV completion is to gauge the $\Uone$ symmetry with $B-L$ charges~\cite{Abe:2020iph, Okada:2020zxo}.
Such a $\Uonebml$ gauge symmetry would be free from gauge anomalies if three right-handed neutrinos are introduced.
Consequently, the WIMP-nucleon amplitude would not  exactly vanish at tree level, but be suppressed by a UV scale, i.e., the breaking scale of the $\Uonebml$ gauge symmetry.
In addition, the pNGB WIMP becomes unstable, and the constraint on its lifetime leads to a UV scale typically exceeding  $\mathcal{O} (10^{11}\text{--}10^{13})~\si{GeV}$~\cite{Abe:2020iph, Okada:2020zxo}.
Such a high scale UV completion can be embedded into a grand unified theory~\cite{Abe:2021byq,
Okada:2021qmi}.

In this work, we would like to decrease the UV scale, because a lower scale may be easier to be probed in future indirect detection and collider experiments.
For this purpose, we assume the pNGB WIMP arising from a \textit{hidden} $\UoneX$ gauge symmetry, where all the SM fields do not carry $\UoneX$ charges.
The gauge anomalies are canceled without introducing right-handed neutrinos, so less fields are involved in this setup.
Since the $\UoneX$ gauge boson does not couple to SM fermions via any $\UoneX$ gauge interaction, the interactions inducing WIMP decays only come from the kinetic mixing between the $\UoneX$ and $\UoneY$ gauge fields, relieving the lifetime constraint on the UV scale.

This paper is organized as follows.
In Section~\ref{sec:model}, we construct a UV-complete model for pNGB WIMP by extending the SM by a hidden $\UoneX$ gauge symmetry and two $\UoneX$-charged scalar fields.
In Section~\ref{sec:pheno}, we discuss the phenomenology of this model regarding the WIMP-nucleon scattering in direct detection experiments, the WIMP lifetime, the related Higgs physics, and the WIMP annihilation relevant to the relic abundance and indirect detection experiments.
Section~\ref{sec:scan} presents results from a random scan in the parameter space.
Section~\ref{sec:sum} gives a summary of the paper.

\section{Model}
\label{sec:model}

We extend the SM with a $\UoneX$ gauge symmetry accompanied with two complex scalar fields $S$ and $\Phi$, which are SM gauge singlets but carry $\UoneX$ charges $q_{S}=1$ and $q_{\Phi}=2$, respectively.
All the SM fields do not carry $\UoneX$ charges.
We assume that $S$ and $\Phi$ develop nonzero vacuum expectation values (VEVs) $v_S$ and $v_\Phi$ with a hierarchy $v_S \ll v_\Phi$.
Thus, $v_\Phi$ represents a UV scale that breaks the $\UoneX$ gauge symmetry into an approximate $\UoneX$ global symmetry.
Beneath the lower scale $v_S$, the $\UoneX$ global symmetry is spontaneously broken, resulting in a pNGB WIMP.

\subsection{Lagrangian}

The $\SUtwoL \times \UoneY \times \UoneX$ gauge-invariant Lagrangian involving $S$, $\Phi$, the SM Higgs doublet $H$, the $\UoneX$ gauge field $X^\mu$, and the $\UoneY$ gauge field $B^\mu$ reads
\begin{eqnarray}\label{eq:Lag}
\mathcal{L} &\supset& (D^{\mu}H)^{\dag}(D_{\mu}H)+(D^{\mu}S)^{\dag}(D_{\mu}S)+(D^{\mu}\Phi)^{\dag}(D_{\mu}\Phi) -\frac{1}{4}B^{\mu\nu}B_{\mu\nu} -\frac{1}{4}X^{\mu\nu}X_{\mu\nu}
\nonumber\\
&&-\frac{s_{\varepsilon}}{2}B^{\mu\nu}X_{\mu\nu}
+\mu^{2}_H |H|^{2}
+\mu_{S}^{2}|S|^{2}
+\mu_{\Phi}^{2}|\Phi|^{2}
-\frac{\lambda_{H}}{2}|H|^{4} -\frac{\lambda_{S}}{2}|S|^{4}
-\frac{\lambda_{\Phi}}{2}|\Phi|^{4}
\nonumber\\
&&-\lambda_{HS}|H|^{2}|S|^{2}
-\lambda_{H\Phi}|H|^{2}|\Phi|^{2}
-\lambda_{S\Phi}|S|^{2}|\Phi|^{2}
+\frac{1}{\sqrt{2}}(\mu_{S\Phi}\Phi^{\dag}S^{2}+\mathrm{H.c.}).
\end{eqnarray}
The covariant derivatives of the scalars are $D_{\mu}H = (\partial_{\mu}-ig' B_{\mu}/2 - ig W_\mu^a \sigma^a/2)H$,
$D_{\mu}S = (\partial_{\mu}-i q_S g_{X}X_{\mu})S$,
and $D_{\mu}\Phi = (\partial_{\mu}-i q_\Phi g_{X}X_{\mu})\Phi$,
where $g_X$ denotes the $\UoneX$ gauge coupling.
The field strengths of $B^\mu$ and $X^\mu$ are defined as $B_{\mu\nu} \equiv \partial_\mu B_\nu - \partial_\nu B_\mu$ and $X_{\mu\nu} \equiv \partial_\mu X_\nu - \partial_\nu X_\mu$.
The $B^{\mu\nu}X_{\mu\nu}$ term implies a kinetic mixing between $B^\mu$ and $X^\mu$ with a mixing parameter $s_\varepsilon \equiv \sin\varepsilon \in (-1,1)$.

The parameter $\mu_{S\Phi}$ can be made real and positive by redefining the phase of $S$, and thus we adopt $\mu_{S\Phi}>0$ hereafter.
The scalar fields can be decomposed as
\begin{eqnarray}
H=\frac{1}{\sqrt{2}}\begin{pmatrix}
0 \\
v+h
\end{pmatrix},\quad
S=\frac{1}{\sqrt{2}}(v_{S}+s+i\eta_{S}),\quad
\Phi=\frac{1}{\sqrt{2}}(v_{\Phi}+\phi+i\eta_{\Phi}),
\end{eqnarray}
where the SM Higgs field is expressed in the unitary gauge and $v = 246.22~\si{GeV}$.
When minimizing the scalar potential, three stationary point conditions are obtained as
\begin{eqnarray}
\label{stationarycondition}
&&\mu_{H}^{2}=\frac{1}{2}(\lambda_{H}v^{2}+\lambda_{HS}v_{S}^{2}+\lambda_{H\Phi}v_{\Phi}^{2}),\nonumber\\
&&\mu_{S}^{2}=\frac{1}{2}(\lambda_{S}v_{S}^{2}+\lambda_{HS}v^{2}+\lambda_{S\Phi}v_{\Phi}^{2})-\mu_{S\Phi}v_{\Phi},\nonumber\\
&&\mu_{\Phi}^{2}=\frac{1}{2}(\lambda_{\Phi}v_{\Phi}^{2}+\lambda_{H\Phi}v^{2}+\lambda_{S\Phi}v_{S}^{2})-\frac{\mu_{S\Phi} v_{S}^{2}}{2v_{\Phi}}.
\end{eqnarray}
The $v_\Phi$ contribution to the $\Phi^\dag S^2$ term leads to
\begin{equation}
\mathcal{L}_\mathrm{soft} = \frac{\mu'^2_S}{4} (S^2 + S^{\dag 2}),
\end{equation}
with $\mu'^2_S = 2\mu_{S\Phi} v_\Phi$.
This is the quadratic term directly introduced in the original pNGB DM model~\cite{Gross:2017dan} to softly break the $\UoneX$ global symmetry.
In the limit $v_\Phi \to \infty$ and $\mu_{S\Phi} \to 0$ with finite $\mu'^2_S$, the original model is recovered.
For a finite $v_\Phi$, there should be some phenomenological deviations from the original model, which will be explored below.

After the scalar fields obtain the nonzero VEVs, the mass terms for the $CP$-even scalars $(h, s, \phi)$ and the $CP$-odd scalars $(\eta_S, \eta_\Phi)$ become
\begin{equation}
\mathcal{L}_{\mathrm{mass}}
	\supset -\frac{1}{2}
\begin{pmatrix}
h & s & \phi
\end{pmatrix}
M_{\mathrm{E}}^{2}
\begin{pmatrix}
h \\
s \\
\phi
\end{pmatrix}
-\frac{1}{2}
\begin{pmatrix}
\eta_{S} & \eta_{\Phi}
\end{pmatrix}
M_{\mathrm{O}}^{2}
\begin{pmatrix}
\eta_{S} \\
\eta_{\Phi} \\
\end{pmatrix},
\end{equation}
where the mass-squared matrices are given by~\cite{Abe:2020iph}
\begin{eqnarray}
\label{mass_e}
M_{\mathrm{E}}^{2} &=& \setlength{\arraycolsep}{0.4em} \begin{pmatrix}
\lambda_{H}v^{2} & \lambda_{HS}vv_{S} & \lambda_{H\Phi}vv_{\Phi} \\
\lambda_{HS}vv_{S} & \lambda_{S}v_{S}^{2} & \lambda_{S\Phi}v_{S}v_{\Phi}-\mu_{S\Phi}v_{S} \\
\lambda_{H\Phi}vv_{\Phi} & \lambda_{S\Phi}v_{S}v_{\Phi}-\mu_{S\Phi}v_{S} & \lambda_{\Phi}v_{\Phi}^{2}+\dfrac{\mu_{S\Phi}v_{S}^{2}}{2v_{\Phi}} 
\end{pmatrix},\\
\label{mass_O}
M_{\mathrm{O}}^{2} &=& \setlength{\arraycolsep}{0.4em}
\mu_{S\Phi}
\begin{pmatrix}
2v_{\Phi} & -v_{S} \\
-v_{S} & \dfrac{v_{S}^{2}}{2v_{\Phi}} \\
\end{pmatrix}.
\end{eqnarray}
The two matrices can be diagonalized by two real orthogonal matrices $U$ and $V$:
\begin{equation}
U^\mathrm{T} M_\mathrm{E}^2 U = \operatorname{diag}(m_{h_1}^2, m_{h_2}^2, m_{h_3}^2),\quad
V^\mathrm{T} M_\mathrm{O}^2 V = \operatorname{diag}(m_{\chi}^2, 0).
\end{equation}
$V$ can be explicitly expressed as
\begin{equation}\setlength{\arraycolsep}{0.4em}
V = \begin{pmatrix}
c_\beta & s_\beta \\
-s_\beta & c_\beta \\
\end{pmatrix},
\end{equation}
with a rotation angle $\beta$ satisfying
\begin{equation}
s_\beta = \frac{v_S}{\sqrt{v_S^2 + 4 v_\Phi^2}}.
\end{equation}
We have adopted the shorthand notations for the trigonometric functions, $s_\beta \equiv \sin\beta$,  $c_\beta \equiv \cos\beta$, and $t_\beta \equiv \tan\beta$.
Such notations are used throughout the paper.

The relations between the interaction and mass bases are
\begin{equation}
\label{UVmatrix}
\begin{pmatrix}
       h \\
       s \\
       \phi \\
\end{pmatrix}
=U\begin{pmatrix}
       h_{1} \\
       h_{2} \\
       h_{3} \\
\end{pmatrix},\quad
\begin{pmatrix}
       \eta_{S} \\
       \eta_{\Phi} \\
\end{pmatrix}
=V\begin{pmatrix}
       \chi \\
       \tilde{\chi} \\
\end{pmatrix}.
\end{equation}
We define $h_{1}$ to be the SM-like Higgs boson with $m_{h_1}=125.10 \pm 0.14~\si{GeV}$~\cite{ParticleDataGroup:2020ssz}, whose dominant component should be $h$, i.e., $|U_{11}| > |U_{21}|, |U_{31}|$.
Similarly, the exotic Higgs bosons $h_2$ and $h_3$ with masses $m_{h_2}$ and $m_{h_3}$ are defined as $s$-like and $\phi$-like, respectively.
We further use positive $U_{11}$, $U_{22}$, and $U_{33}$ to fix the signs of the $U$ matrix elements.
$\tilde{\chi}$ is the massless Nambu-Goldstone boson associated with the spontaneous breaking of the $\UoneX$ gauge symmetry, while $\chi$ is a pNGB WIMP with a mass squared of
\begin{eqnarray}
\label{eq:muc}
m_{\chi}^{2}=\frac{\mu_{S\Phi}}{2v_{\Phi}} (v_{S}^{2}+4v_{\Phi}^{2}),
\end{eqnarray}
serving as a DM candidate.
The typical range for $m_\chi$ would be $\mathcal{O}(\si{GeV})\text{--}\mathcal{O}(\si{TeV})$.

If $\mu_{S\Phi} = 0$, the Lagrangian \eqref{eq:Lag} respects two distinct $\Uone$ global symmetries, one with $S \to e^{i\alpha_1} S$ and the other one with $\Phi \to e^{i\alpha_2} \Phi$.
Consequently, both $\eta_{S}$ and $\eta_{\Phi}$ are massless Nambu-Goldstone bosons according to the Goldstone theorem~\cite{Nambu:1960tm, Goldstone:1961eq}.
Nonetheless, the existence of the $\mu_{S\Phi}$ term merges the two $\Uone$ symmetries into the  $\UoneX$ global symmetry with $q_{S}=1$ and $q_{\Phi}=2$.
As a result, only $\tilde\chi$ remains massless, while $\chi$ obtains a mass proportional to $\sqrt{\mu_{S\Phi}}$.

After the spontaneous breaking of the $\SUtwoL \times \UoneY \times \UoneX$ gauge symmetry, the gauge fields obtain the mass terms
\begin{equation}\setlength{\arraycolsep}{0.4em}
\mathcal{L}_{\mathrm{mass}}
\supset m_{W}^{2}W^{-,\mu}W_{\mu}^{+} + \frac{1}{2}
\begin{pmatrix}
B^\mu & W^{3,\mu} & X^{\mu}
\end{pmatrix}
M_\mathrm{N}^2
\begin{pmatrix}
B_{\mu} \\
W^{3}_{\mu} \\
X_{\mu} \\
\end{pmatrix}, 
\end{equation}
with $m_{W}^{2} = g^{2}v^{2}/4$ and
\begin{equation}\setlength{\arraycolsep}{0.4em}
M_\mathrm{N}^2= \begin{pmatrix}
	g'^{2}v^{2}/4 & -gg'v^{2}/4 & 0 \\
	-gg'v^{2}/4  & g^{2}v^{2}/4 & 0 \\
	0 & 0 & g_{X}^{2}(v_{S}^{2}+4v_{\Phi}^{2}) \\
\end{pmatrix}.
\end{equation}
Considering both such a $B_\mu$-$W^3_\mu$ mass mixing and the $B_\mu$-$X_\mu$ kinetic mixing,
the physical neutral gauge fields $(A_\mu, Z_\mu, Z'_\mu)$ can be obtained through a linear transformation~\cite{Babu:1997st, Lao:2020inc}
\begin{equation}
\begin{pmatrix}
B_{\mu} \\
W^{3}_{\mu} \\
X_{\mu} \\
\end{pmatrix}
= K
\begin{pmatrix}
A_{\mu} \\
Z_{\mu} \\
Z'_{\mu} \\
\end{pmatrix}
\end{equation}
with
\begin{equation}\setlength{\arraycolsep}{0.4em}
K = \begin{pmatrix}
\hat{c}_\mathrm{W} & - \hat{s}_\mathrm{W} c_\xi - t_\varepsilon s_\xi & \hat{s}_\mathrm{W} s_\xi - t_\varepsilon c_\xi \\
\hat{s}_\mathrm{W} & \hat{c}_\mathrm{W} c_\xi & - \hat{c}_\mathrm{W} s_\xi \\
0 & s_\xi/c_\varepsilon & c_\xi/c_\varepsilon \\
\end{pmatrix}.
\end{equation}
Here we denote $\hat{s}_\mathrm{W} \equiv \sin\hat{\theta}_W$ and $\hat{c}_\mathrm{W} \equiv \cos\hat{\theta}_W$, where $\hat{\theta}_\mathrm{W} \equiv \tan^{-1} (g'/g)$ is the weak mixing angle.
$\varepsilon$ is the angle related to the kinetic mixing, and $\xi$ is a rotation angle determined by the equation~\cite{Babu:1997st}
\begin{equation}
t_{2 \xi}=
\frac{s_{2\varepsilon}\hat{s}_\mathrm{W} v^{2}(g^{2}+g^{\prime 2})}
{c_{\varepsilon}^{2} v^{2}(g^{2}+g^{\prime 2})(1-\hat{s}_\mathrm{W}^{2} t_{\varepsilon}^{2})-4 g_{X}^{2} (v_{S}^{2}+4v_{\Phi}^{2})}.
\end{equation}

The gauge fields $(A_\mu, Z_\mu, Z'_\mu)$ have canonical kinetic terms as well as diagonalized mass terms.
The corresponding masses for the photon, $Z$, and $Z'$ bosons are given by $m_A = 0$ and~\cite{Chun:2010ve}
\begin{equation}\label{m_z_zp}
m_{Z}^{2}=\frac{v^{2}}{4}(g^{2}+g'^{2})(1+\hat{s}_\mathrm{W}t_{\varepsilon}t_{\xi}),\quad
m_{Z'}^{2}=\frac{g_{X}^{2}(v_{S}^{2}+4v_{\Phi}^{2})}{c_{\varepsilon}^{2}(1+\hat{s}_\mathrm{W}t_{\varepsilon}t_{\xi})},
\end{equation}
respectively.
Define $r \equiv m_{Z'}^2/m_Z^2$, and we can further derive~\cite{Lao:2020inc}
\begin{equation}
t_{\xi}=\frac{2\hat{s}_\mathrm{W}t_{\varepsilon}}{1-r}\left[1+\sqrt{1-r\left(\frac{2\hat{s}_\mathrm{W}t_{\varepsilon}}{1-r}\right)^2}\,\right]^{-1}.
\end{equation}
$r \gg 1$ would lead to
\begin{equation}\label{eq:t_xi}
t_{\xi} \simeq - \frac{\hat{s}_{\mathrm{W}} t_{\varepsilon}}{r} +\mathcal{O}
(r^{- 2}).
\end{equation}
If there is no kinetic mixing between $B_\mu$ and $X_\mu$, we have $\varepsilon = \xi = 0$, $m_{Z}^{2} = (g^{2}+g'^{2})v^{2}/4$, and $m_{Z'}^{2}= g_{X}^{2}(v_{S}^{2}+4v_{\Phi}^{2})$.

\subsection{Interactions}

In the basis of the mass eigenstates $\chi$ and $h_i$, the scalar trilinear couplings can be expressed as
\begin{eqnarray}
\label{app:potential}
\mathcal{L}_{\mathrm{tri}} = - \frac{1}{2} \sum_{i=1}^{3}g_{h_{i}\chi^2}h_{i}\chi^2
- \sum_{i,j,k=1}^{3}g_{h_{i}h_{j}h_{k}}h_{i}h_{j}h_{k},
\end{eqnarray}
where
\begin{eqnarray}
g_{h_{i}\chi^2} &=& (\lambda_{HS}c_\beta^2+\lambda_{H\Phi}s_\beta^2)vU_{1i}+(\lambda_{S} v_{S} c_\beta^2 +\lambda_{S\Phi} v_{S} s_\beta^2 + 2\mu_{S\Phi} s_\beta c_\beta)U_{2i}\nonumber \\
&& +[\lambda_{\Phi} v_{\Phi} s_\beta^2 +(\lambda_{S\Phi} v_{\Phi} + \mu_{S\Phi}) c_\beta^{2}]U_{3i},
\\
g_{h_{i}h_{j}h_{k}}&=& \frac{1}{2}(\lambda_{H}vU_{1i}+\lambda_{HS}v_{S}U_{2i}+\lambda_{H\Phi}v_{\Phi}U_{3i})U_{1j}U_{1k}\nonumber \\
&& +\frac{1}{2}[\lambda_{HS}vU_{1i}+\lambda_{S}v_{S}U_{2i}+(\lambda_{S\Phi}v_{\Phi}-\mu_{S\Phi})U_{3i}]U_{2j}U_{2k}\nonumber \\
&& +\frac{1}{2}(\lambda_{H\Phi}vU_{1i}+\lambda_{S\Phi}v_{S}U_{2i}+\lambda_{\Phi}v_{\Phi}U_{3i})U_{3j}U_{3k}.
\end{eqnarray}
The Yukawa couplings become
\begin{equation}
\mathcal{L}_{h_i ff} =  - \sum\limits_f \sum\limits_{i = 1}^3 \frac{m_f U_{1i}}{v}\, h_i\bar ff ,                   
\end{equation}
where $f$ denotes the SM fermions.

The neutral current interactions arising from the $\SUtwoL \times \UoneY \times \UoneX$ gauge symmetry are given by
\begin{equation}
\mathcal{L}_{\mathrm{NC}} = B_\mu j_{\mathrm{Y}}^\mu  + W_\mu ^3j_3^\mu  + X_\mu j_X^\mu,
\end{equation}
with
\begin{eqnarray}
j_{\mathrm{Y}}^\mu  &=& \sum\limits_f g'(Y_{f,\mathrm{L}}{\bar f}_{\mathrm{L}}\gamma ^\mu f_{\mathrm{L}} + Y_{f,\mathrm{R}}{\bar f}_{\mathrm{R}}\gamma ^\mu f_{\mathrm{R}}) ,
\\
j_3^\mu  &=& \sum\limits_f gT_f^3{\bar f}_{\mathrm{L}}\gamma ^\mu f_{\mathrm{L}} ,
\\
j_X^\mu  &=& i g_X(S^\dag \overleftrightarrow {\partial ^\mu }S + 2 \Phi ^\dag \overleftrightarrow {\partial ^\mu }\Phi ),
\end{eqnarray}
where $T_f^3$ is the third component of the weak isospin for a SM fermion $f$, and $Y_{f,\mathrm{L}}$ and $Y_{f,\mathrm{R}}$ are the weak hypercharges for left- and right-handed fermions.
If we define $\hat{A}_\mu \equiv \hat{c}_\mathrm{W} B_\mu + \hat{s}_\mathrm{W} W_\mu^3$ and $\hat{Z}_\mu \equiv - \hat{s}_\mathrm{W} B_\mu + \hat{c}_\mathrm{W} W_\mu^3$, the neutral current interactions can be expressed in a familiar form
\begin{equation}
\mathcal{L}_{\mathrm{NC}} = {\hat A}_\mu j_{\mathrm{EM}}^\mu  + {\hat Z}_\mu {\hat j}_Z^\mu  + X_\mu j_X^\mu,
\end{equation}
with
\begin{eqnarray}
j_{\mathrm{EM}}^\mu  &=& \sum\limits_f Q_f e\bar f\gamma ^\mu f,
\\
\hat{j}_Z^\mu &=& \frac{g}{2{\hat c}_{\mathrm{W}}}\sum\limits_f \bar f{\gamma ^\mu }(T_f^3 - 2Q_f\hat s_{\mathrm{W}}^2 - T_f^3\gamma ^5)f, 
\end{eqnarray}
where $Q_f$ is the electric charge of $f$ in units of $e = gg'/\sqrt{g^2 + g'^2}$.

The relation between $(\hat{A}_\mu, \hat{Z}_\mu, X_\mu)$ and the mass basis $(A_\mu, Z_\mu, Z'_\mu)$ is
\begin{equation}
\begin{pmatrix}
   {\hat A}_\mu   \\
   {\hat Z}_\mu   \\
   X_\mu   \\
\end{pmatrix} = R\begin{pmatrix}
   A_\mu   \\
   Z_\mu   \\
   Z'_\mu   \\
\end{pmatrix},
\end{equation}
where
\begin{equation}\setlength{\arraycolsep}{0.4em}
R = \begin{pmatrix}
   {\hat c}_{\mathrm{W}} & {\hat s}_{\mathrm{W}} & 0  \\
    - {\hat s}_{\mathrm{W}} & {\hat c}_{\mathrm{W}} & 0  \\
   0 & 0 & 1  \\
\end{pmatrix}K = \begin{pmatrix}
   1 &  - {\hat c}_{\mathrm{W}} t_\varepsilon s_\xi &  - {\hat c}_{\mathrm{W}} t_\varepsilon c_\xi \\
   0 & {\hat s}_{\mathrm{W}} t_\varepsilon s_\xi + c_\xi  & {\hat s}_{\mathrm{W}} t_\varepsilon c_\xi - s_\xi   \\
   0 & s_\xi /c_\varepsilon  & c_\xi /c_\varepsilon   \\
\end{pmatrix}.
\end{equation}
Therefore, the neutral current interactions in the mass basis are~\cite{Frandsen:2011cg,
Lao:2020inc}
\begin{equation}
\mathcal{L}_{\mathrm{NC}} = A_\mu j_{\mathrm{EM}}^\mu  + Z_\mu j_Z^\mu  + Z'_\mu j_{Z'}^\mu ,
\end{equation}
with
\begin{eqnarray}
j_Z^\mu  &=& \sum\limits_f \bar f\gamma ^\mu (g_{\mathrm{V},Z}^f - g_{\mathrm{A},Z}^f\gamma ^5)f  + \frac{s_\xi }{c_\varepsilon }j_X^\mu ,
\\
j_{Z'}^\mu  &=& \sum\limits_f \bar f\gamma ^\mu (g_{\mathrm{V},Z'}^f - g_{\mathrm{A},Z'}^f\gamma ^5)f  + \frac{c_\xi }{c_\varepsilon }j_X^\mu ,
\end{eqnarray}
where
\begin{eqnarray}
g_{\mathrm{V},Z}^f &=& \frac{g}{2{\hat c}_{\mathrm{W}}}({\hat s}_{\mathrm{W}}t_\varepsilon s_\xi  + c_\xi )(T_f^3 - 2Q_f\hat s_{\mathrm{W}}^2) - Q_f e{\hat c}_{\mathrm{W}}t_\varepsilon s_\xi ,
\\
g_{\mathrm{A},Z}^f &=& \frac{g}{2{\hat c}_{\mathrm{W}}}({\hat s}_{\mathrm{W}}t_\varepsilon s_\xi  + c_\xi )T_f^3,
\\
g_{\mathrm{V},Z'}^f &=& \frac{g}{2{{\hat c}_{\mathrm{W}}}}({\hat s}_{\mathrm{W}}t_\varepsilon c_\xi  - s_\xi )(T_f^3 - 2Q_f\hat s_{\mathrm{W}}^2) - Q_f e{\hat c}_{\mathrm{W}}t_\varepsilon c_\xi ,
\\
g_{\mathrm{A},Z'}^f &=& \frac{g}{2{\hat c}_{\mathrm{W}}}({\hat s}_{\mathrm{W}}t_\varepsilon c_\xi  - s_\xi )T_f^3.
\end{eqnarray}
Note that the electromagnetic current interactions $A_\mu j_\mathrm{EM}^\mu$ remain in the SM form.
For $\varepsilon = 0$, we have $A^\mu = \hat{A}^\mu$, $Z^\mu = \hat{Z}^\mu$, and $Z'^\mu = X^\mu$, and the $Z$ couplings to the fermions are the same as in the SM, while the $Z'$ boson only couples to $j^\mu_X$.
The existence of the kinetic mixing makes $Z$ couple to $j^\mu_X$ and $Z'$ couple to the SM fermions.

The $Z$-$\chi$-$h_i$ and $Z'$-$\chi$-$h_i$ couplings from the neutral current interactions $Z_\mu j_Z^\mu  + Z'_\mu j_{Z'}^\mu$ are
\begin{equation}\label{eq:coupling:Z_Zp_chi_hi}
\mathcal{L}_{\chi h_i} = \sum\limits_{i=1}^{3} (g_{Z\chi h_i}Z_\mu \chi \overleftrightarrow {\partial ^\mu} h_i + g_{Z'\chi h_i}Z'_\mu \chi \overleftrightarrow {\partial ^\mu} h_i),
\end{equation}
where
\begin{eqnarray}
g_{Z\chi h_i} &=& \frac{g_X s_\xi }{c_\varepsilon }(c_\beta U_{2i} - 2 s_\beta U_{3i}),
\\
g_{Z'\chi h_i} &=& \frac{ g_X c_\xi }{c_\varepsilon} (c_\beta U_{2i} - 2 s_\beta U_{3i}).
\end{eqnarray}
These couplings break the $Z_2$ symmetry $\chi \to -\chi$,  inducing decay processes of the pNGB WIMP $\chi$.
In order to be a viable DM candidate, $\chi$ should have a sufficiently long lifetime.

The best measured electroweak quantities are the fine-structure constant $\alpha(m_Z)$ in the $\overline{\mathrm{MS}}$ scheme, the Fermi constant $G_\mathrm{F}$, and the $Z$ boson pole mass $m_Z$.
From these quantities, it is conventional to define the ``physical'' weak mixing parameters $s_\mathrm{W}^2$ and $c_\mathrm{W}^2 \equiv 1 - s_\mathrm{W}^2$ via the tree-level SM relation~\cite{Babu:1997st, Burgess:1993vc}
\begin{equation}
s_\mathrm{W}^2 c_\mathrm{W}^2 = \frac{\pi\alpha}{\sqrt{2} G_\mathrm{F} m_Z^2}.
\end{equation}
Then the hatted parameters $\hat{s}_\mathrm{W}^2$ and $\hat{c}_\mathrm{W}^2$ can be determined by~\cite{Chun:2010ve}
\begin{eqnarray}
\label{solve_swh}
&&s_\mathrm{W}^{2}c_\mathrm{W}^{2}(1+\hat{s}_\mathrm{W}t_{\varepsilon}t_{\xi})=\hat{s}_\mathrm{W}^2 \hat{c}_\mathrm{W}^2.
\end{eqnarray}
The values of $g$ and $g'$ are settled by $g = e/\hat{s}_\mathrm{W}$ and $g' = e/\hat{c}_\mathrm{W}$ with $e = \sqrt{4\pi\alpha}$.
 
There are 10 free parameters in the model, which are chosen to be 
\begin{eqnarray}
\label{11para}
v_{S},~ v_{\Phi},~ m_{\chi},~ m_{h_{2}},~ m_{h_{3}},~ m_{Z'},~ \lambda_{HS},\ \lambda_{H\Phi},~ \lambda_{S\Phi},~ s_{\varepsilon}.
\end{eqnarray}
For a UV completion of the original pNGB DM model, we are particularly interested in the parameter regions with $v_\Phi \gg v_S \sim v$.
This implies a mass hierarchy of $m_{Z'} \sim m_{h_3} \gg m_{h_2} \sim m_{h_1}$, leading to $r = m_{Z'}^2/m_Z^2 \gg 1$ and hence $|\xi| \ll |\varepsilon|$.
Therefore, the value of $\hat{s}_\mathrm{W}$ would be very close to $s_\mathrm{W}$ for any value of $s_\varepsilon$.

The kinetic mixing contributes to the electroweak oblique parameters $S$, $T$, and $U$~\cite{Peskin:1990zt,Peskin:1991sw} at tree level.
For a small $\varepsilon$, we have
\begin{equation}
\xi \simeq \frac{s_{\mathrm{W}} \varepsilon}{1 - r},
\end{equation}
and~\cite{Holdom:1990xp}
\begin{equation}
S \simeq \frac{4 (c^2_{\mathrm{W}} - r) s^2_{\mathrm{W}} c^2_{\mathrm{W}}
\varepsilon^2}{\alpha (1 - r)^2},\quad
T \simeq - \frac{r s^2_{\mathrm{W}} \varepsilon^2}{\alpha (1 - r)^2},\quad
U \simeq \frac{4 s^4_{\mathrm{W}} c^2_{\mathrm{W}} \varepsilon^2}{\alpha (1 - r)^2}.
\end{equation}
Because $S$, $T$, and $U$ are all highly suppressed by $r$, we expect that the constraint on the oblique parameters from the global fit of electroweak precision measurements~\cite{Baak:2014ora} would not constrain our interested parameter regions.

\section{Phenomenology}
\label{sec:pheno}

In this section, we discuss the phenomenological consequences of the model.

\subsection{WIMP-nucleon Scattering}

In the original pNGB DM model with a directly introduced soft breaking parameter $\mu'^2_S$, the WIMP-nucleon scattering amplitude exactly vanishes at tree level in the zero momentum transfer limit~\cite{Gross:2017dan}.
Our UV completion gives $\mu'^2_S$ a dynamical origin, but inevitably introduces the $\chi$-$\chi$-$\phi$ coupling, leading to a nonvanishing $\chi$-nucleon scattering amplitude.
Nonetheless, we expect that the amplitude is significantly suppressed by a high UV scale $v_\Phi$, since the $v_\Phi \to \infty$ limit recovers the original model.

The spin-independent (SI) $\chi$-nucleon scattering is induced by the $\chi$-quark scattering via $t$-channel exchanges of the $CP$-even Higgs bosons $h_1$, $h_2$, and $h_3$.
In the zero momentum transfer limit, the tree-level $\chi$-quark scattering amplitude becomes
\begin{eqnarray}
i\mathcal{M} &=& \frac{i m_{q}}{v}\,\bar{u}(k_{2})u(k_{1})
\left(
\frac{g_{h_{1}\chi^{2}}U_{11}}{m_{h_{1}}^{2}}
+\frac{g_{h_{2}\chi^{2}}U_{12}}{m_{h_{2}}^{2}}
+\frac{g_{h_{3}\chi^{2}}U_{13}}{m_{h_{3}}^{2}}
\right),
\nonumber\\
&=& \frac{i m_{q}}{v}\,\bar{u}(k_{2})u(k_{1})
\begin{pmatrix}
g_{h_{1}\chi^{2}} & g_{h_{2}\chi^{2}} & g_{h_{3}\chi^{2}}
\end{pmatrix}
\begin{pmatrix}
m_{h_{1}}^{-2} & & \\
 & m_{h_{2}}^{-2} & \\
 & & m_{h_{3}}^{-2} \\
\end{pmatrix}
U^\mathrm{T}
\begin{pmatrix}
1 \\
0 \\
0 \\
\end{pmatrix},
\end{eqnarray}
where $u(k_{1})$ and $\bar{u}(k_{2})$ denote the plane-wave spinor coefficients for the incoming and outgoing quarks $q$ of 4-momenta $k_1$ and $k_2$.
It is equivalent~\cite{Gross:2017dan, Jiang:2019soj} to express the amplitude in the interaction basis $(h, s, \phi)$, whose couplings to $\chi$ are given by $G = \begin{pmatrix} g_{h\chi^2} & g_{s\chi^2} & g_{\phi\chi^2} \end{pmatrix}$
with
\begin{eqnarray}
g_{h\chi^2} &=& (\lambda_{HS}c_\beta^2+\lambda_{H\Phi}s_\beta^{2})v,
\\
g_{s\chi^2} &=& \lambda_{S}v_{S} c_\beta^2 + \lambda_{S\Phi}v_{S} s_\beta^2 + 2\mu_{S\Phi}s_\beta c_\beta,
\\
g_{\phi\chi^2} &=& \lambda_{\Phi}v_{\Phi}s_\beta^2+(\lambda_{S\Phi}v_{\Phi}+\mu_{S\Phi})c_\beta^2.
\end{eqnarray}
Utilizing $\operatorname{diag}(m_{h_{1}}^{-2},~ m_{h_{2}}^{-2},~ m_{h_{3}}^{-2}) = U^\mathrm{T} (M_\mathrm{E}^{2})^{-1} U$ and $\begin{pmatrix}g_{h_{1}\chi^{2}} & g_{h_{2}\chi^{2}} & g_{h_{3}\chi^{2}}\end{pmatrix} = GU$, we derive
\begin{eqnarray}
i\mathcal{M} = \frac{i m_{q}}{v}\,\bar{u}(k_{2})u(k_{1})
G (M_\mathrm{E}^{2})^{-1}
\begin{pmatrix}
1 \\
0 \\
0 \\
\end{pmatrix}.
\end{eqnarray}

Expanding the scattering amplitude in orders of $v_\Phi$ for $v_\Phi \gg v, v_S$, we obtain
\begin{equation}
i\mathcal{M} \simeq \frac{i \tilde{\lambda} m_q m_\chi ^2}{2v^2 v_\Phi ^2}\, \bar u(k_2)u(k_1) + \mathcal{O}(v_\Phi^{-4}),
\end{equation}
where
\begin{equation}
\tilde{\lambda} = \frac{\lambda _{H\Phi }\lambda _{S\Phi } - \lambda _\Phi \lambda _{HS} + 2\lambda _{HS}\lambda _{S\Phi } - 2\lambda _S \lambda _{H\Phi }}{\lambda _H\lambda _S\lambda _\Phi  + 2\lambda _{HS}\lambda _{H\Phi }\lambda _{S\Phi } - \lambda _S\lambda _{H\Phi }^2 - \lambda _\Phi \lambda _{HS}^2 - \lambda _H\lambda _{S\Phi }^2}.
\end{equation}
Thus, the amplitude is suppressed by $v_\Phi^{-2}$, as expected.
Based on effective field theory~\cite{Yu:2011by}, we derive the resulting SI $\chi$-nucleon scattering cross section
\begin{equation}\label{eq:sigma_SI}
\sigma_{\chi N}^\mathrm{SI} \simeq \frac{\tilde{\lambda}^2 m_N^4 m_\chi^4 [2 + 7(f_u^N + f_d^N + f_s^N)]^2}{1296\pi (m_N + m_\chi)^2 v^4 v_\Phi^4} + \mathcal{O}(v_\Phi^{-6}),
\end{equation}
which is suppressed by $v_\Phi^{-4}$.
Here, $f_{u,d,s}^{N}$ are the nucleon form factors for light quarks~\cite{Ellis:2000ds}.

\begin{figure}[!t]
\centering
\subfigure[$m_\chi$ vs. $\sigma_{\chi N}^\mathrm{SI}$\label{fig:ddchi}]
{\includegraphics[width=0.48\textwidth]{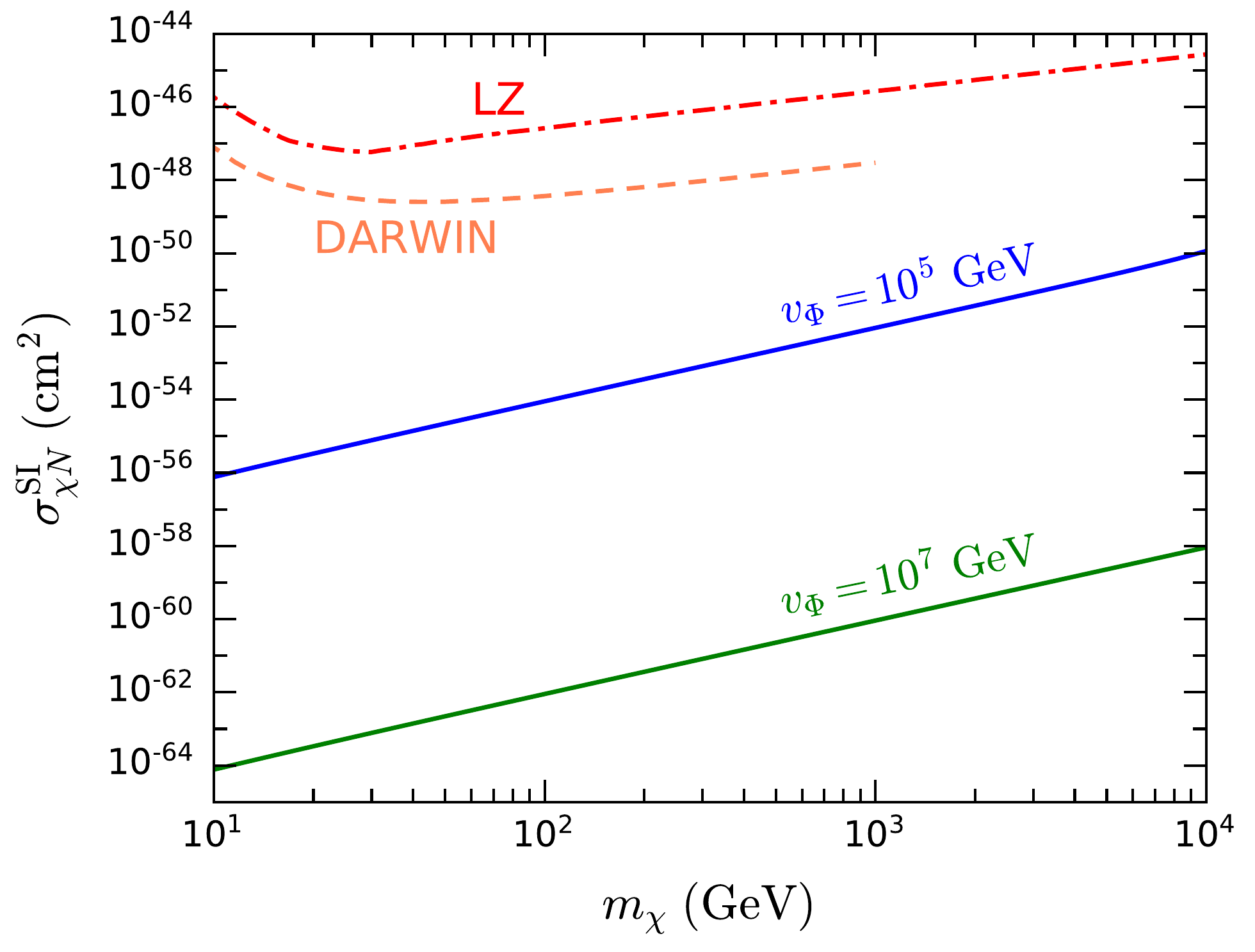}}
\hspace{.01\textwidth}
\subfigure[$v_\Phi$ vs. $\sigma_{\chi N}^\mathrm{SI}$\label{fig:ddvph}]
{\includegraphics[width=0.48\textwidth]{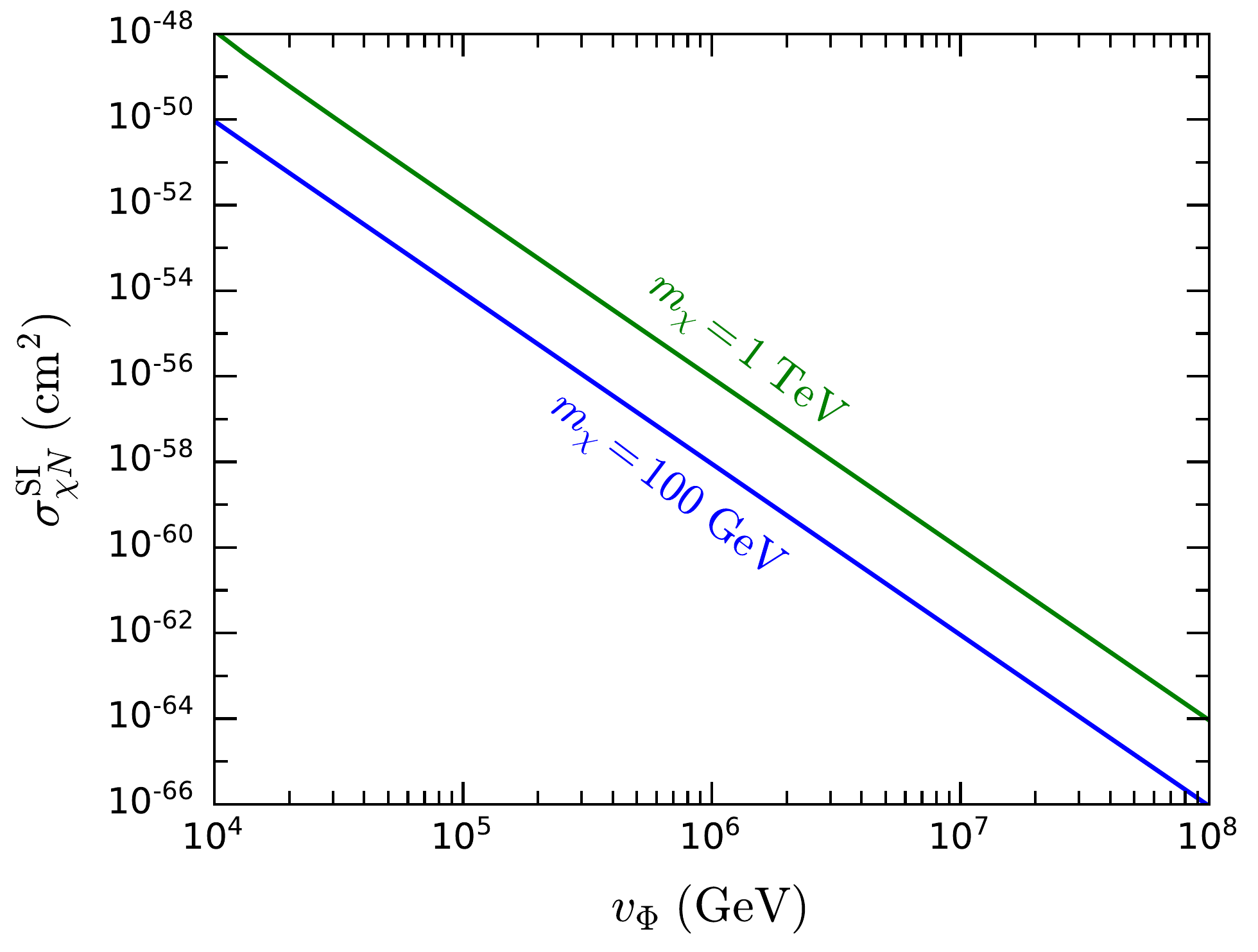}}
\caption{SI $\chi$-nucleon scattering cross section $\sigma_{\chi N}^\mathrm{SI}$ as functions of $m_\chi$ (a) and $v_\Phi$ (b).
The blue and green lines in the left (right) panel correspond to $v_\Phi = 10^{5}~\si{GeV}$ ($m_\chi = 100~\si{GeV}$) and $v_\Phi = 10^{7}~\si{GeV}$ ($m_\chi = 1~\si{TeV}$), respectively.
Other parameters are fixed as $v_{S}=1~\si{TeV}$, $m_{h_{2}}=300~\si{GeV}$, $m_{h_{3}}=0.1 v_{\Phi}$, $\lambda_{HS}=0.03$, and $\lambda_{H\Phi} = \lambda_{S\Phi}=0.01$.
The left panel also shows the 90\% C.L. constraint from the LZ direct detection experiment~\cite{Aalbers:2022fxq} as well as the future sensitivity of the DARWIN experiment~\cite{DARWIN:2016hyl}.}
\label{fig:dd}
\end{figure}

In Fig.~\ref{fig:ddchi}, we plot the $\chi$-nucleon scattering cross section $\sigma_{\chi N}^\mathrm{SI}$ as functions of $m_\chi$ for $v_\Phi = 10^{5}$ and $10^{7}~\si{GeV}$, with the other related parameters fixed to be $v_{S}=1~\si{TeV}$, $m_{h_{2}}=300~\si{GeV}$, $m_{h_{3}}=0.1 v_{\Phi}$, $\lambda_{HS}=0.03$, and $\lambda_{H\Phi} = \lambda_{S\Phi}=0.01$.
For $m_\chi \gg m_N$, Eq.~\eqref{eq:sigma_SI} shows that $\sigma_{\chi N}^\mathrm{SI}$ is proportional to $m_\chi^2$.
Therefore, as $m_\chi$ increases by one order of magnitudes, $\sigma_{\chi N}^\mathrm{SI}$ in Fig.~\ref{fig:ddchi} increases by two orders of magnitudes.
$v_\Phi = 10^{7}~\si{GeV}$ leads to cross sections smaller that those for $v_\Phi = 10^{5}~\si{GeV}$ by eight orders of magnitudes, because of  $\sigma_{\chi N}^\mathrm{SI} \propto v_\Phi^{-4}$.
Note that $v_\Phi = 10^{5}~\si{GeV}$ results in $\sigma_{\chi N}^\mathrm{SI}$ much smaller than the 90\% confidence level (C.L.) upper limits from the recent LZ direct detection experiment~\cite{Aalbers:2022fxq}, and even beyond the reach of the future DARWIN experiment with a $200~\si{t\cdot yr}$ exposure~\cite{DARWIN:2016hyl}.
Fig.~\ref{fig:ddvph} displays $\sigma_{\chi N}^\mathrm{SI}$ as functions of $v_{\Phi}$ for $m_\chi = 100~\si{GeV}$ and $1~\si{TeV}$, demonstrating an obvious $\sigma_{\chi N}^\mathrm{SI} \propto v_\Phi^{-4}$ behavior.

\subsection{WIMP Lifetime}

\begin{figure}[!t]
\centering
\subfigure[~$\chi \to h_i^{(*)} Z^{(*)}$]
{\includegraphics[width=0.3\textwidth]{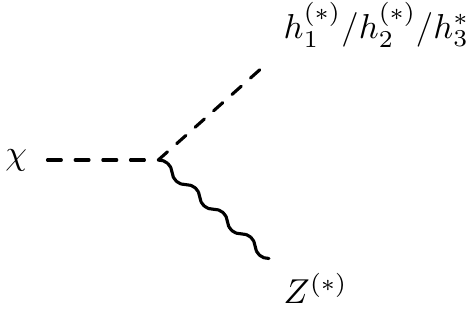}}
\hspace{2em}
\subfigure[~$\chi \to h_i^{(*)} Z'^*$]
{\includegraphics[width=0.3\textwidth]{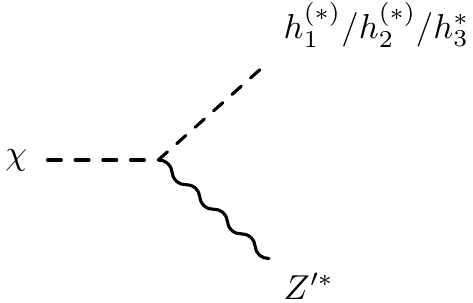}}
\caption{Feynman diagrams for $\chi$ decays into $h_i^{(*)} Z^{(*)}$ (a) and $h_i^{(*)} Z'^*$ (b), where $h_1$, $h_2$, and $Z$ could be either on or off shell, while $h_3$ and $Z'$ must be off shell for $m_\chi \ll m_{h_3}, m_{Z'}$.}
\label{fig:chi_decay}
\end{figure}

The $Z$-$\chi$-$h_i$ and $Z'$-$\chi$-$h_i$ couplings \eqref{eq:coupling:Z_Zp_chi_hi} induce the decay of the pNGB WIMP $\chi$.
We are particularly interested in the parameter regions with $m_\chi \ll m_{Z'}\sim m_{h_3}$, where the $\chi$ decay processes involve $\chi \to h_i^{(*)} Z^{(*)} $ and $\chi \to h_i^{(*)} Z'^*$. Depending on the mass spectrum, the $h_1$, $h_2$, and $Z$ bosons could be either on or off shell, while the $h_3$ and $Z'$ bosons must be off shell.
The corresponding Feynman diagrams are depicted in Fig.~\ref{fig:chi_decay}.

Six years of Fermi-LAT $\gamma$-ray observations of dwarf galaxies imply a conservative constraint on the WIMP lifetime, $\tau_{\chi} \gtrsim 10^{27}~\si{s}$~\cite{Baring:2015sza}, corresponding to a bound on the total WIMP decay width, $\Gamma_\chi \equiv 1/\tau_{\chi} \lesssim  \num{6.6e-52}~\si{GeV}$. 
In the $v_\Phi \to \infty$ limit, all the $\chi$ decay channels are forbidden, and $\chi$ becomes stable.
Thus, the constraint on the $\chi$ lifetime is expected to give a lower bound on the UV scale $v_\Phi$.
Therefore, the total decay width of the pNGB WIMP $\chi$ should be carefully calculated.

When $m_\chi > m_{h_i} + m_Z$ ($i=1,2$), the 2-body partial decay width of $\chi \to h_{i} Z$ is given by
\begin{equation}
\Gamma (\chi \to h_{i}Z) = \frac{g_{Z\chi h_{i}}^2 m_\chi^3}{16\pi m_Z^2} \;\lambda^{3/2}\left(1,\frac{m_{h_{i}}^2}{m_{\chi}^2},\frac{m_Z^2}{m_{\chi}^2}\right),
\end{equation}
where the $\lambda$ function is defined as $\lambda(x,y,z)=x^{2}+y^{2}+z^{2}-2xy-2xz-2yz$.
When $m_\chi > m_{h_i} + 2m_f$ ($i=1,2$), we should consider the 3-body decays $\chi \to h_i f \bar{f}$.
If $m_{h_i} + 2m_f < m_\chi < m_{h_i} + m_Z$, both the Feynman diagrams mediated by the off-shell $Z$ and $Z'$ bosons contribute to $\chi \to h_i f \bar{f}$.
However, once the 2-body decay channels $\chi \to h_{i} Z$ open, the $\chi \to h_i f \bar{f}$ decay diagrams mediated by the $Z$ bosons should be discarded for avoiding double counting.
When $m_Z + 2m_f < m_\chi < m_{h_i} + m_Z$ ($i=1,2$), 
the 3-body decays $\chi \to Z f \bar{f}$ mediated by $h_i$ should be involved.
Since the fermion couplings to $h_i$ are commonly suppressed by $m_f/v$, the dominant contributions to $\chi \to Z f \bar{f}$ come from the heaviest SM fermions $t$, $b$, $\tau$, and $c$.
If $2m_W + m_Z < m_\chi < m_{h_2} + m_Z$, the 3-body decays $\chi \to W^+W^-Z$ and $\chi \to ZZZ$ mediated by off-shell $h_i$ bosons may happen.
Nonetheless, our calculation shows that their contributions are negligible, compared to $\chi \to Z f \bar{f}$ and $\chi \to h_i f \bar{f}$.
If all 2- and 3-body decay channels are kinematically forbidden, the 4-body decays $\chi \to f\bar{f}f'\bar{f}'$ should be taken into account.

For calculating the 3-body partial decay widths, we derive analytic expressions and perform numerical integrals.
Since the Feynman diagrams and integrals for the 4-body decays are too complicated to be dealt with by hand, we utilize the Monte Carlo tool \texttt{MadGraph5\_aMG@NLO}~\cite{Alwall:2014hca} to automatically evaluate the 4-body partial decay widths.
In the latter approach, \texttt{FeynRules}~\cite{Alloul:2013bka} is used to implement the model.

\begin{figure}[!t]
\centering
\subfigure[$m_\chi$ vs. $\Gamma_\chi$\label{fig:mchiDW}]
{\includegraphics[width=0.48\textwidth]{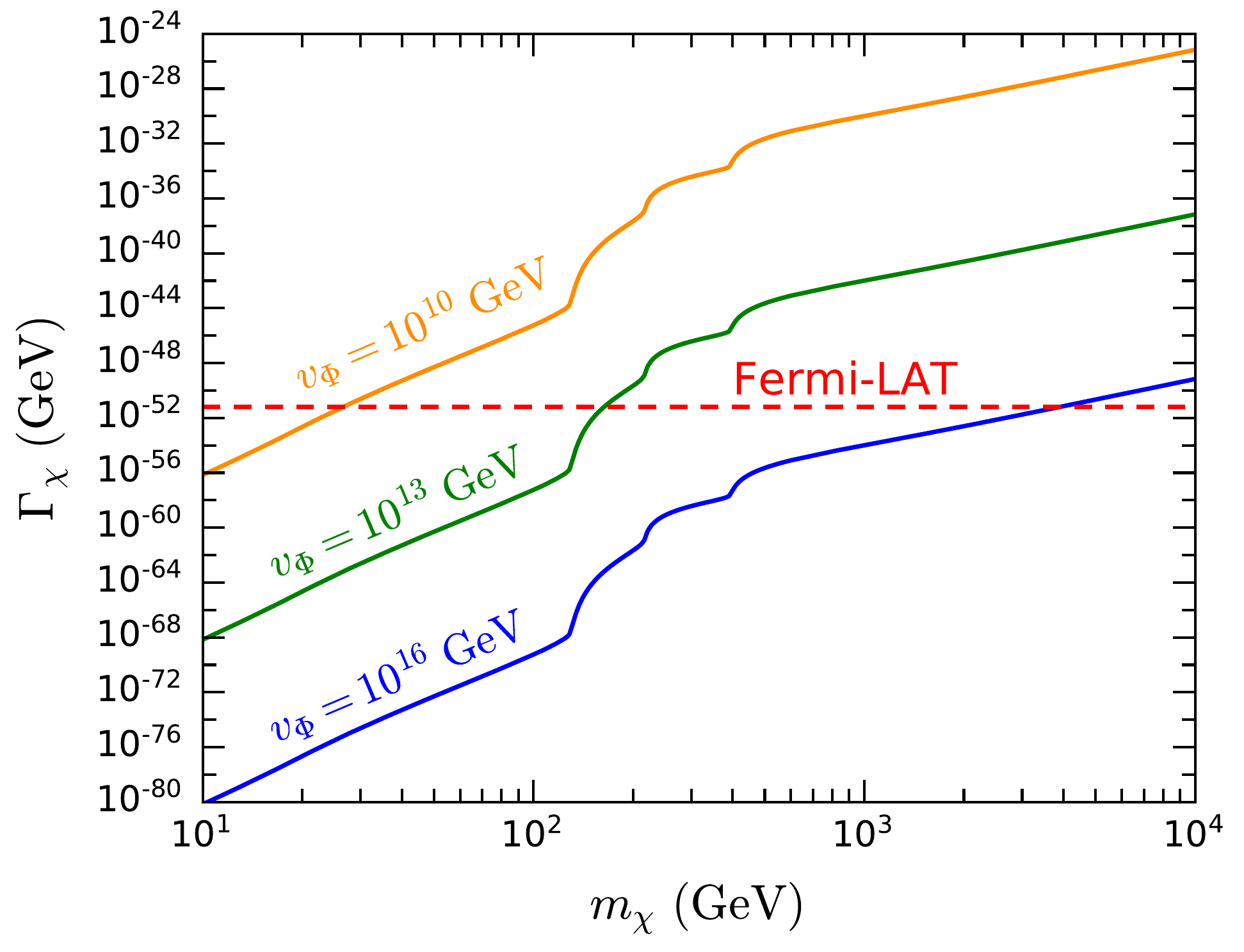}}
\hspace{.01\textwidth}
\subfigure[$v_\Phi$ vs. $\tau_\chi$\label{fig:vphDW}]
{\includegraphics[width=0.48\textwidth]{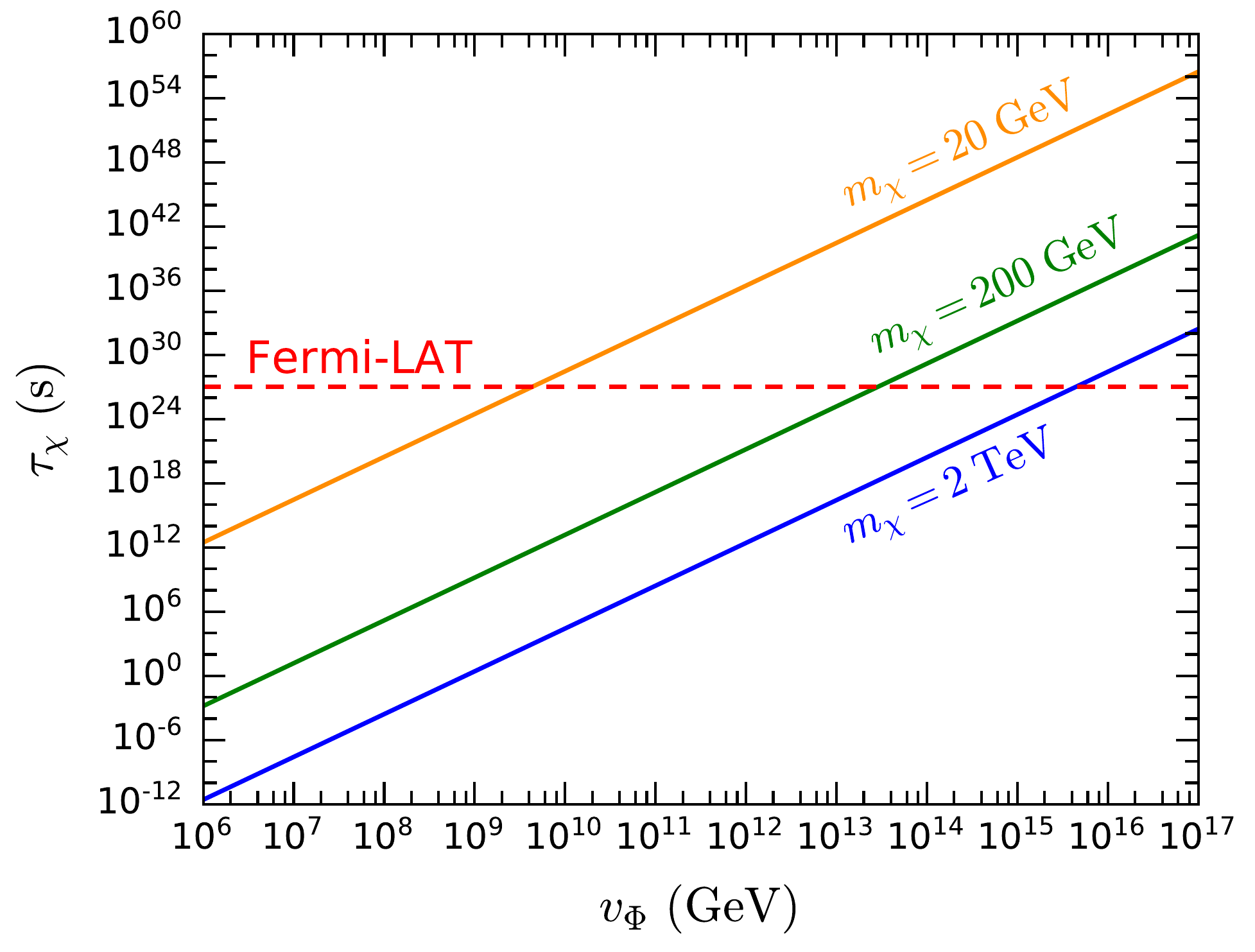}}
\caption{$\chi$ decay width $\Gamma_\chi$ as functions of $m_\chi$ for $v_\Phi = 10^{10}$, $10^{13}$, and $10^{15}~\si{GeV}$ (a) and $\chi$ lifetime $\tau_{\chi}$ as functions of $v_\Phi$ for $m_\chi = 20$, $200$, and $2000~\si{GeV}$ (b).
Other parameters are fixed as $v_{S}=1~\si{TeV}$, $m_{h_{2}}=300~\si{GeV}$, $m_{h_{3}}=m_{Z'}=0.1v_\Phi$, $\lambda_{HS}=0.03$,  $\lambda_{H\Phi} = \lambda_{S\Phi}=0.01$, and  $\sin\varepsilon=0.1$.
The dashed lines denote the conservative constraint $\tau_{\chi} \gtrsim 10^{27}~\si{s}$ from Fermi-LAT $\gamma$-ray observations of dwarf galaxies~\cite{Baring:2015sza}.}
\label{fig:longevity}
\end{figure}

We fix the parameters as $v_{S}=1~\si{TeV}$, $m_{h_{2}}=300~\si{GeV}$, $m_{h_{2}}=m_{Z'}=0.1v_\Phi$, $\lambda_{HS}=0.03$,  $\lambda_{H\Phi} = \lambda_{S\Phi}=0.01$, and  $\sin\varepsilon=0.1$, and show the total decay width of $\chi$ as functions of $m_\chi$ for $v_\Phi = 10^{10}$, $10^{13}$, and $10^{15}~\si{GeV}$ in Fig.~\ref{fig:mchiDW}.
The threshold effects at $m_\chi = m_{h_1} \simeq 125~\si{GeV}$, $m_\chi = m_{h_1} + m_Z \simeq 216~\si{GeV}$, and $m_\chi = m_{h_2} + m_Z \simeq 391~\si{GeV}$ are clearly demonstrated.
The red dashed line denotes the Fermi-LAT bound on the WIMP lifetime~\cite{Baring:2015sza} converted to the total decay width, $\Gamma_\chi \lesssim  \num{6.6e-52}~\si{GeV}$.
Thus, $m_\chi \gtrsim 25$, $140$, and $3900~\si{GeV}$ are excluded for $v_\Phi = 10^{10}$, $10^{13}$, and $10^{15}~\si{GeV}$, respectively.
In Fig.~\ref{fig:vphDW}, we display the $\chi$ lifetime as functions of $v_\Phi$ for $m_\chi = 20$, $200$, and $2000~\si{GeV}$, where
the Fermi-LAT constraint excludes $v_\Phi \lesssim \num{4e9}$, $\num{3e13}$, and $\num{4e15}~\si{GeV}$, respectively. 
Thus, the bound on $v_\Phi$ given by the WIMP lifetime 
is much more stringent than the bound from direct detection experiments.

\subsection{Higgs Physics}

In this model, the properties of the SM-like Higgs boson $h_1$ deviate from the SM prediction.
The tree-level $h_1$ couplings to SM particles can be parametrized as
\begin{equation}
\mathcal{L}_{h_{1}}
= \kappa_{W}\, \frac{2m_W^2}{v}\, h_{1}W_{\mu}^{+}W^{-,\mu} + \kappa _Z\,\frac{m_Z^2}{v}\, h_{1}Z_{\mu}Z^{\mu} - \sum_{f}\kappa_{f}\,\frac{m_{f}}{v}\, h_{1}\bar{f}f,
\end{equation}
where $\kappa_{W}$, $\kappa_{Z}$, and $\kappa_{f}$ are the modifiers to the couplings with $W$, $Z$, and fermions.
The SM corresponds to $\kappa_{W} = \kappa_{Z} = \kappa_{f} = 1$, while this model gives
\begin{eqnarray}
\kappa_{W} &=& \kappa_f = U_{11},
\label{k_W}\\
\kappa_{Z} &=& U_{11} c_\xi ^2(1 + \hat{s}_\mathrm{W} t_\varepsilon  t_\xi )  + \frac{s_\xi ^2 g_X^2 v}{c_\varepsilon ^2 m_Z^2}(U_{21} v_S + 4 U_{31} v_\Phi ).
\label{k_Z}
\end{eqnarray}

In addition, exotic Higgs decay channels may exist.
If $m_{h_1} > 2 m_\chi$, the invisible decay channel $h_{1} \to \chi\chi$ opens, leading to an invisible decay width
\begin{equation}
\Gamma(h_{1} \to \chi\chi)=\frac{g_{h_{1}\chi^2}^2}{32\pi m_{h_1}}\sqrt{1-\frac{4m_{\chi}^2}{m_{h_1}^2}}.
\end{equation}
If $m_{h_1} > m_\chi + m_Z$, there is a semi-invisible decay channel $h_{1} \to \chi Z$ with a partial decay width
\begin{equation}
\Gamma(h_{1} \to \chi Z)=\frac{g_{Z\chi h_{1}}^2 m_{h_1}^3}{16\pi m_Z^2} \; \lambda^{3/2}\left(1,\frac{m_\chi^2}{m_{h_1}^2},\frac{m_Z^2}{m_{h_1}^2}\right),
\end{equation}
If $m_{h_1} > 2 m_{h_2}$, the partial decay width of $h_{1} \to h_{2}h_{2}$ is given by
\begin{equation}
\Gamma(h_{1} \to  h_{2}h_{2})=\frac{(g_{h_1 h_2 h_2}+g_{h_2 h_1 h_2}+g_{h_2 h_2 h_1})^2}{8\pi m_{h_1}}\sqrt{1-\frac{4m_{h_2}^2}{m_{h_1}^2}}.
\end{equation}

We utilize a numerical tool \texttt{Lilith~2}~\cite{Bernon:2015hsa, Kraml:2019sis} to constrain the model parameter space with the LHC Higgs measurements based on the \texttt{Lilith} database of version 19.09, including ATLAS and CMS Run 2 data of integrated luminosity $36~\mathrm{fb}^{-1}$. The important results sensitive to this model come from the measurements of $h_1\to \gamma\gamma$~\cite{ATLAS:2018hxb}, $h_1\to ZZ$~\cite{ATLAS:2017azn}, and $h_1\to W^+W^-$~\cite{ATLAS:2018xbv}, and the search for invisible Higgs decays~\cite{CMS:2018yfx}, and  the combined measurements of the Higgs couplings from several channels~\cite{CMS:2018uag}.
For each parameter point, \texttt{Lilith} constructs an approximate likelihood function from the measurements of the Higgs signal strengths.
The corresponding $p$-value larger than $0.05$ is required, ensuring that each viable parameter point is consistent with the experimental results at $95\%$ C.L.

\subsection{WIMP Annihilation}

The relic abundance of the pNGB WIMP $\chi$ is essentially determined by $\chi\chi$ annihilation cross section at the freeze-out epoch, $\langle\sigma_{\mathrm{ann}}v\rangle_{\mathrm{FO}}$.
Potential $\chi\chi$ annihilation channels include $f\bar{f}$, $W^+ W^-$, $ZZ$, and $h_i h_j$ ($i,j=1,2$).
The related Feynman diagrams are enormous.
We make use of a \texttt{MadGraph5\_aMG@NLO} plugin \texttt{MadDM}~\cite{Ambrogi:2018jqj} to automatically generate and calculate all tree-level annihilation diagrams, and to solve the Boltzmann equation for predicting the  relic abundance $\Omega_{\chi}h^{2}$.

$\chi\chi$ annihilation would still occur at the present day, inducing potential $\gamma$-ray signals in indirect detection experiments.
A combined search for such $\gamma$-ray signals in the dwarf galaxies from the Fermi-LAT space experiment and the MAGIC Cherenkov telescopes~\cite{MAGIC:2016xys} have given important constraints on the DM annihilation cross section.
We further utilize \texttt{MadGraph5\_aMG@NLO}~\cite{Alwall:2014hca} to evaluate the $\chi\chi$ annihilation cross section at a typical average WIMP velocity $\num{2e-5}$ for dwarf galaxies, $\langle\sigma_{\mathrm{ann}}v\rangle_{\mathrm{D}}$.
Thus, the Fermi-MAGIC result can be used to constrain the model.

\section{Parameter Scan}
\label{sec:scan}

We perform a random scan in the following parameter ranges,
\begin{eqnarray}
&&10\ \mathrm{GeV}< v_{S}, m_{h_2}, m_\chi <10^{4}\ \mathrm{GeV},\quad 10^{9}\ \mathrm{GeV}<v_{\Phi}<10^{17}\ \mathrm{GeV},\nonumber\\
&& 10^{8}\ \mathrm{GeV}<m_{h_{3}},\ m_{Z'}<10^{16}\ \mathrm{GeV},\quad 0.01<|s_\varepsilon|<0.9,\nonumber\\
&& 10^{-3}<|\lambda_{HS}|,\ |\lambda_{H\Phi}|,\ |\lambda_{S\Phi}|<1.
\end{eqnarray}
The induced couplings $\lambda_H$,  $\lambda_S$,  $\lambda_\Phi$, and $g_X$ are further required to range from $10^{-3}$ to $1$.
We select the parameter points that satisfy the phenomenological requirements below.
\begin{itemize}
\item The WIMP lifetime satisfies the Fermi-LAT bound $\tau_{\chi} \gtrsim 10^{27}~\si{s}$~\cite{Baring:2015sza}.
\item The signal strengths of the 125~GeV Higgs boson $h_1$ are consistent with the Higgs measurements after LHC Run~2 at 95\% C.L. according to the \texttt{Lilith} calculation.
\item The predicted WIMP relic abundance $\Omega_\chi h^2$ lies within the $3\sigma$ range of the Planck measurement $\Omega_\mathrm{DM} h^2 = 0.1200\pm 0.0012$~\cite{Planck:2018vyg}.
\end{itemize}

\begin{figure}[!t]
\centering
\subfigure[$v_S$-$m_{h_2}$ plane\label{fig:vs_mh2}]
{\includegraphics[width=0.48\textwidth]{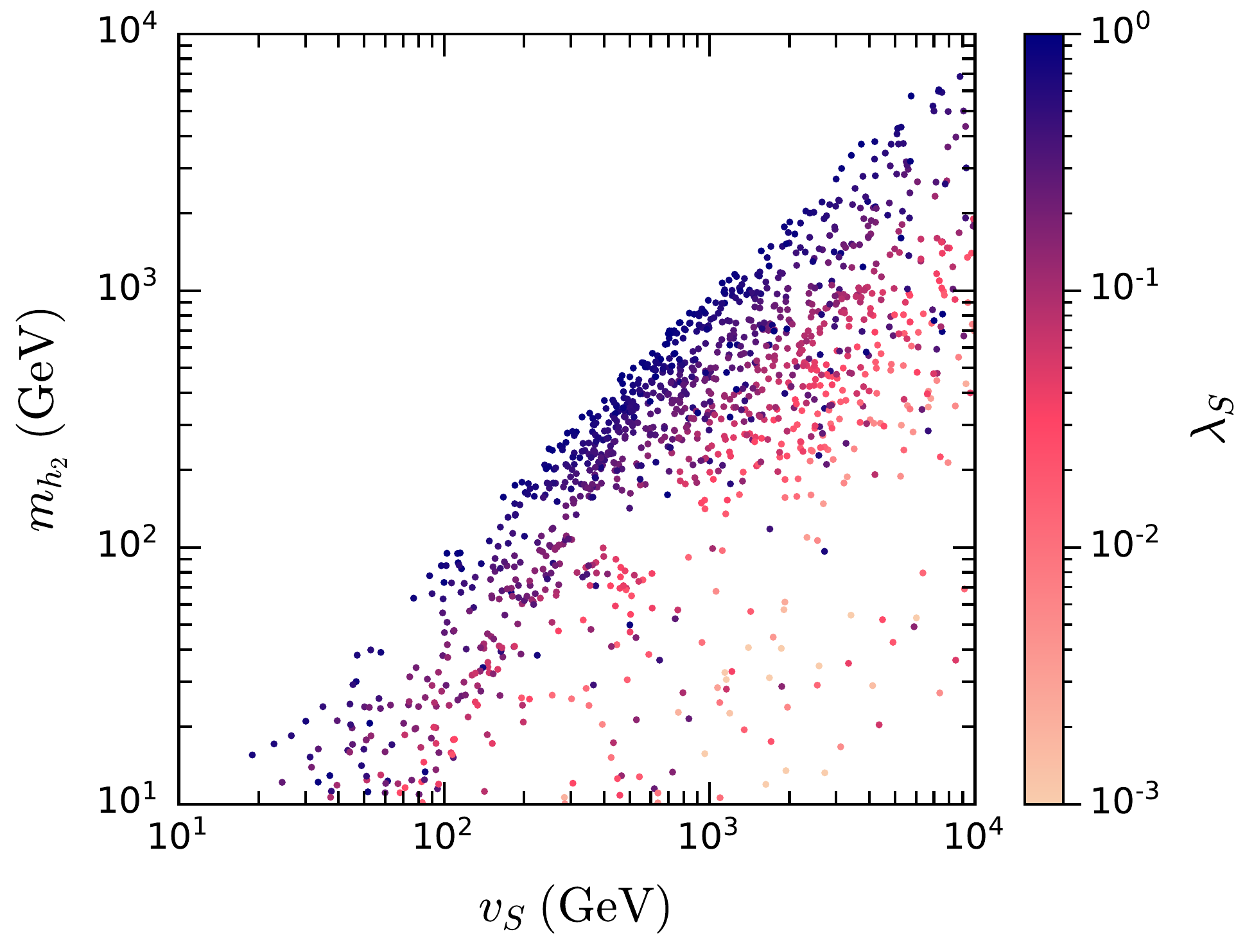}}
\hspace{.01\textwidth}
\subfigure[$v_\Phi$-$m_{h_3}$ plane\label{fig:vph_mh3}]
{\includegraphics[width=0.48\textwidth]{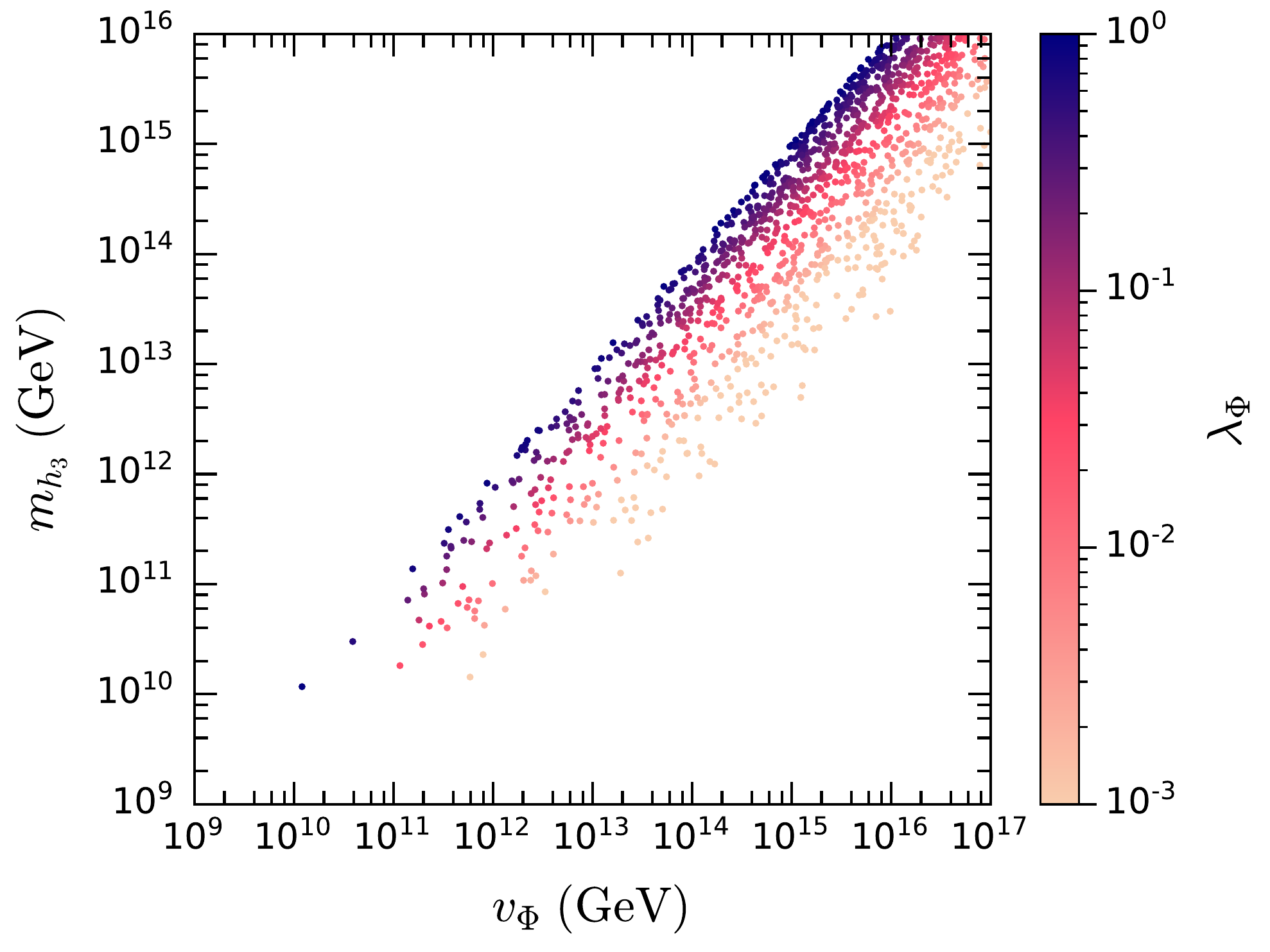}}
\subfigure[$v_\Phi$-$m_{Z'}$ plane\label{fig:vph_mzp}]
{\includegraphics[width=0.48\textwidth]{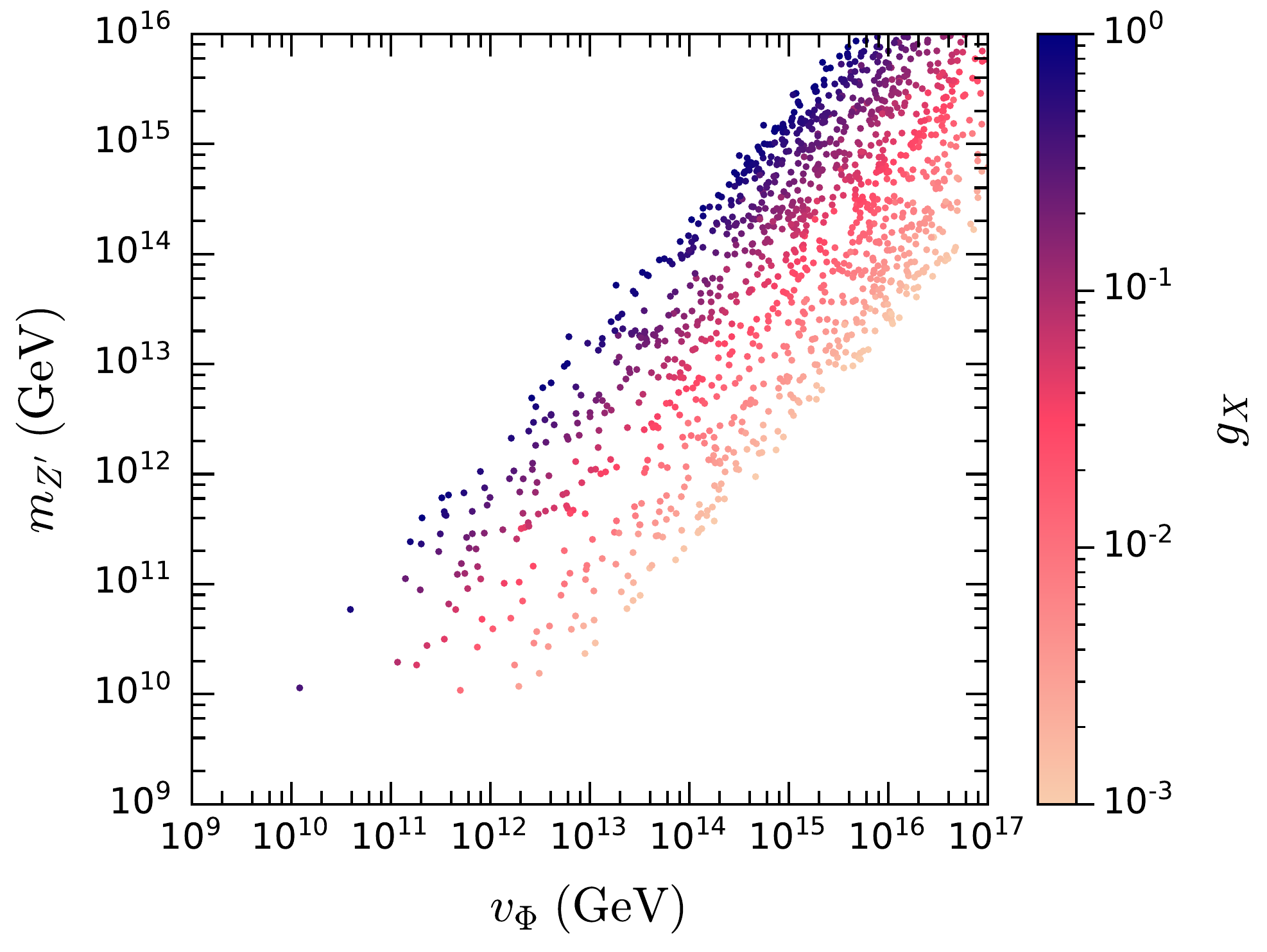}}
\hspace{.01\textwidth}
\subfigure[$m_{Z'}$-$|\xi|$ plane\label{fig:mzp_xi}]
{\includegraphics[width=0.48\textwidth]{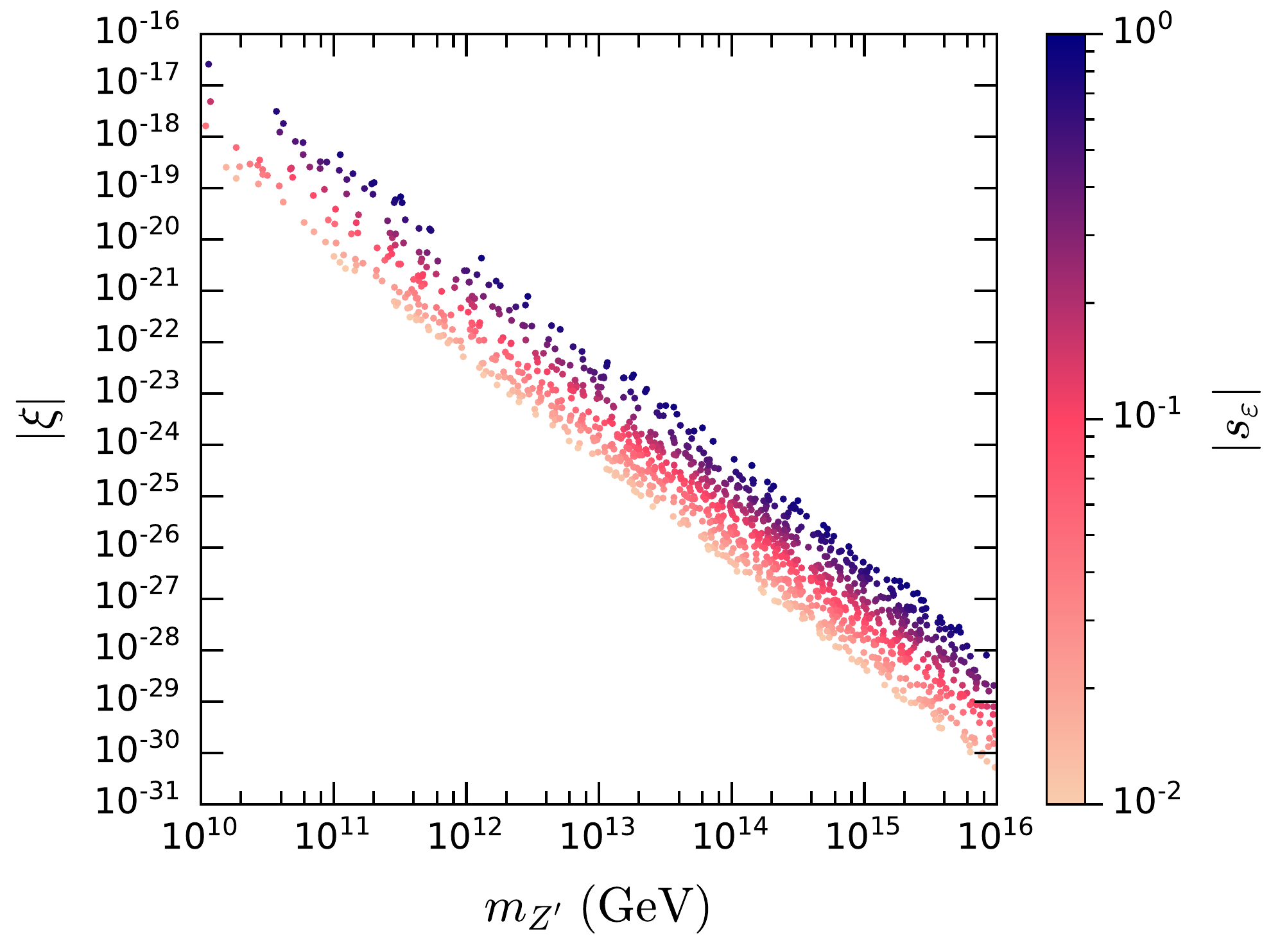}}
\caption{Selected parameter points projected onto the $v_S$-$m_{h_2}$ (a), $v_\Phi$-$m_{h_3}$ (b), $v_\Phi$-$m_{Z'}$ (c), and $m_{Z'}$-$|\xi|$ (d) planes, with color axes corresponding to $\lambda_S$, $\lambda_\Phi$, $g_X$, and $|s_\varepsilon|$, respectively.}
\end{figure}

We project the selected parameter points onto the $v_S$-$m_{h_2}$, $v_\Phi$-$m_{h_3}$, $v_\Phi$-$m_{Z'}$, and $m_{Z'}$-$|\xi|$ planes in Figs.~\ref{fig:vs_mh2}, \ref{fig:vph_mh3}, \ref{fig:vph_mzp}, and \ref{fig:mzp_xi}, with color axes corresponding to $\lambda_S$, $\lambda_\Phi$, $g_X$, and $|s_\varepsilon|$, respectively.
Since we focus on the parameter region with $v_\Phi \gg v,v_S,m_\chi$, the mass-squared matrix \eqref{mass_e} implies $m_{h_3}^2 \simeq \lambda_\Phi v_\Phi^2$, which is clearly shown in Fig.~\ref{fig:vph_mh3}.
More precisely, this plot demonstrates that $m_{h_3}$ is proportional to $v_\Phi$ and positively correlated to $\lambda_\Phi$.
For $|\lambda_{HS}| \ll 1$, Eq.~\eqref{mass_e} leads to $m_{h_2}^2 \simeq \lambda_S v_S^2$.
Nonetheless, such positive correlations of $m_{h_2}$ to $v_S$ and $\lambda_S$ do not totally manifest in Fig.~\ref{fig:vs_mh2}.
The exceptions should be due to large $|\lambda_{HS}|$.

According to Eqs.~\eqref{m_z_zp} and \eqref{eq:t_xi}, $v_\Phi, m_{Z'} \gg v,v_S$ means that $m_{Z'} \simeq 2g_X v_\Phi/c_\varepsilon$ and $t_\xi \simeq - s_\mathrm{W} t_\varepsilon m_Z^2/m_{Z'}^2$.
Fig.~\ref{fig:vph_mzp} illustrates the positive correlations of $m_{Z'}$ to $v_\Phi$ and $g_X$, while Fig.~\ref{fig:mzp_xi} displays the negative (positive) correlation of $|\xi|$ to $m_{Z'}$ ($|s_\varepsilon|$).
From Figs.~\ref{fig:vph_mh3} and \ref{fig:vph_mzp}, we find that the lower limit of the UV scale $v_\Phi$ is down to $\sim 10^{10}~\si{GeV}$, given by the Fermi-LAT constraint on the WIMP lifetime.

\begin{figure}[!t]
\centering
\subfigure[$\Gamma_{h_1}$-$(1-\kappa_Z)$ plane\label{fig:Gammah1_kappaZ}]
{\includegraphics[width=0.48\textwidth]{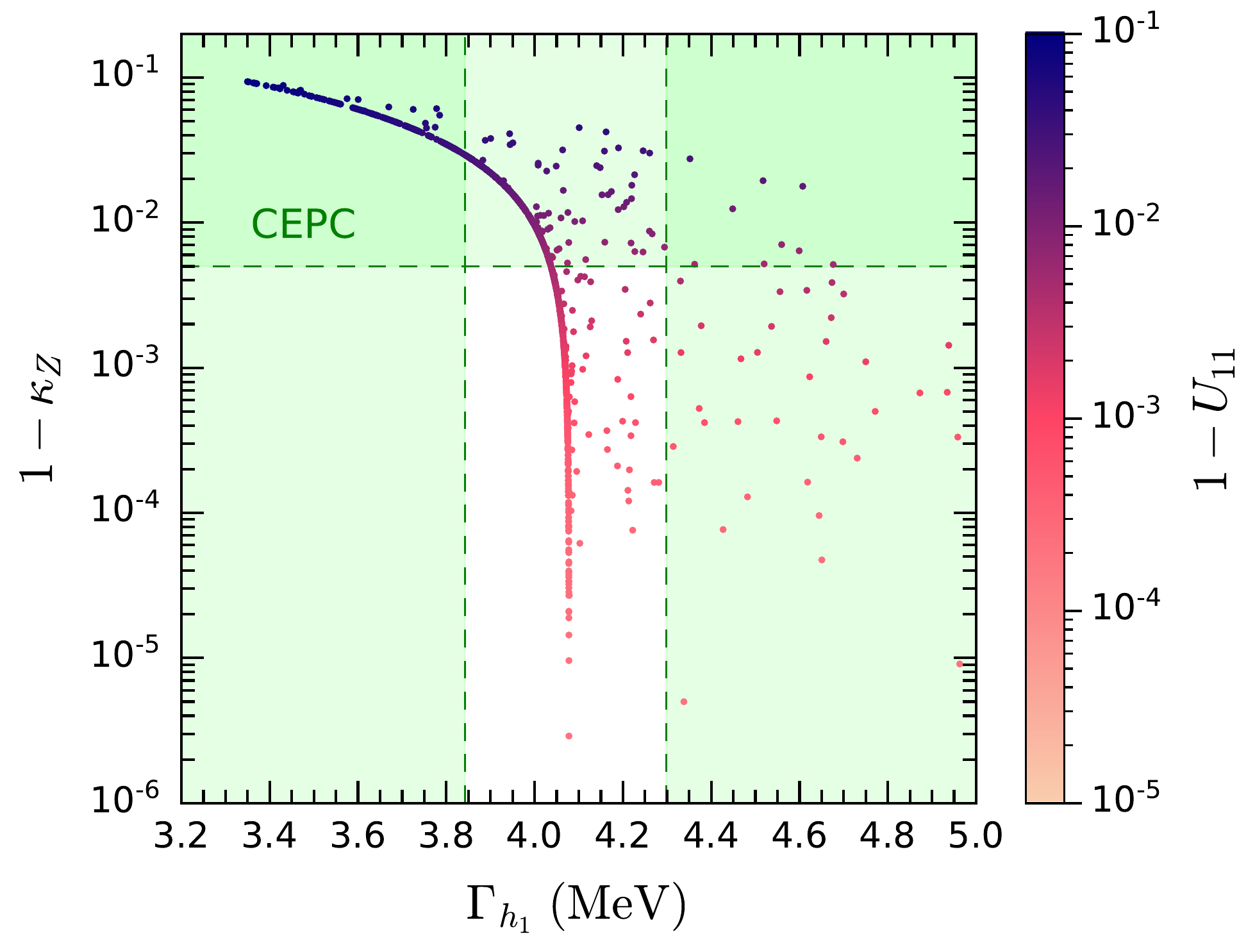}}
\hspace{.01\textwidth}
\subfigure[$m_\chi$-$\mathrm{BR}_\mathrm{inv}$ plane\label{fig:mchi_BRinv}]
{\includegraphics[width=0.48\textwidth]{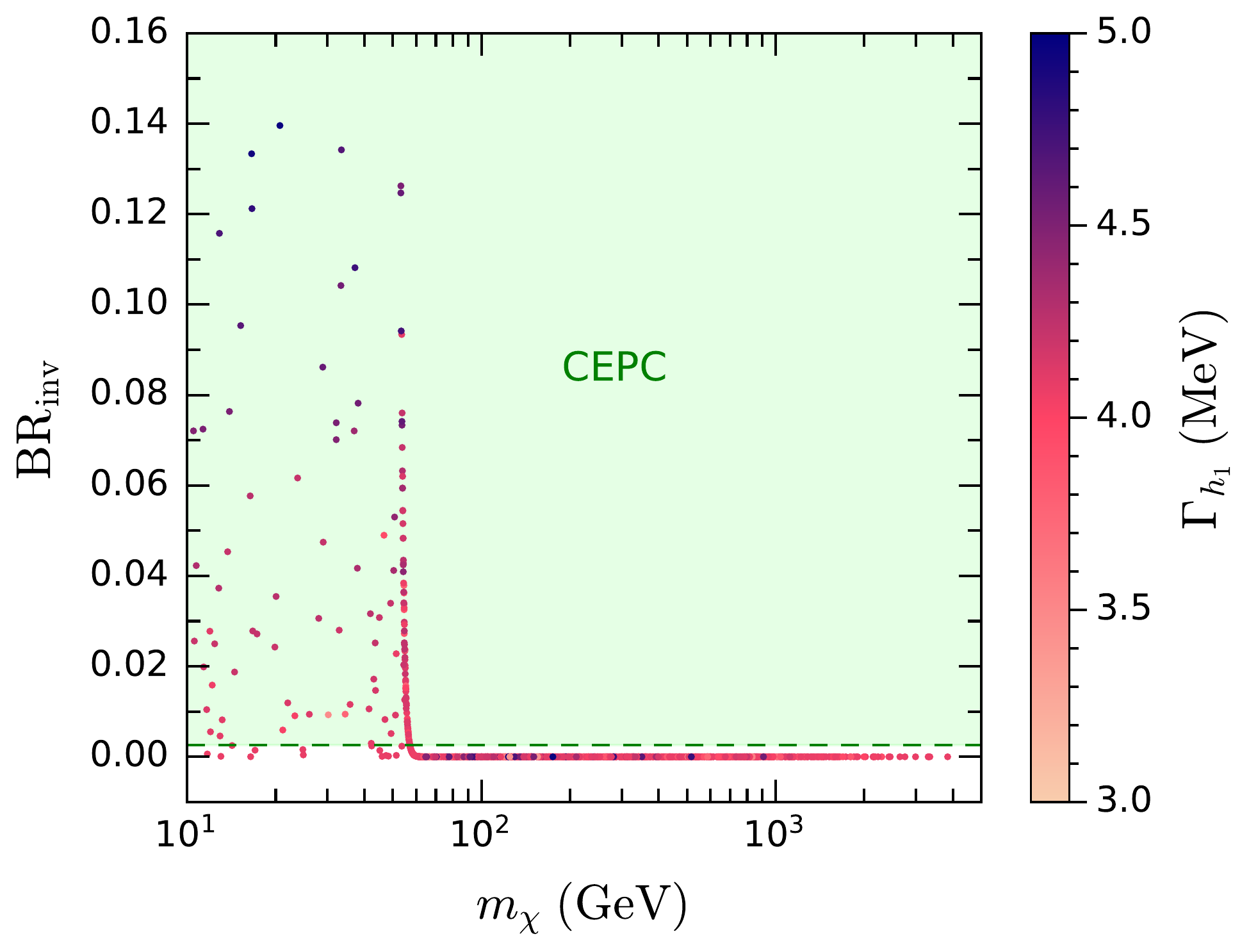}}
\caption{Selected parameter points projected onto the $\Gamma_{h_1}$-$(1-\kappa_Z)$ (a) and $m_\chi$-$\mathrm{BR}_\mathrm{inv}$ (b) planes, with color axes corresponding to $1-U_{11}$ and $\Gamma_{h_1}$, respectively.
The green regions denote the expected coverage of the future CEPC experiment at $95\%$ C.L.~\cite{CEPCStudyGroup:2018ghi}.}
\end{figure}

In Fig.~\ref{fig:Gammah1_kappaZ}, we demonstrate the selected parameter points in the $\Gamma_{h_1}$-$(1-\kappa_Z)$ plane, with colors denoting $1 - U_{11}$.
For $v_\Phi, m_{Z'} \gg v,v_S$, Eq.~\eqref{k_Z} becomes $\kappa_Z \simeq U_{11} = \kappa_W = \kappa_f$, and thus the color axis is basically identical to the vertical axis in  Fig.~\ref{fig:Gammah1_kappaZ}.
There is an obvious curve constituted by parameter points in this plot, indicating the positive correlation between the $h_1$ total decay width $\Gamma_{h_1}$ and $\kappa_Z$ (or $U_{11}$, equivalently).
The exceptional parameter points have larger $\Gamma_{h_1}$, which are contributed by the exotic Higgs decays $h_1 \to \chi\chi$, $h_1 \to \chi Z$, and $h_1 \to h_2 h_2$.
We can see that the constraints from current LHC Higgs measurements give $1-\kappa_Z \lesssim 0.1$ (or $U_{11} \gtrsim 0.9$) and $3.3~\si{MeV} \lesssim\Gamma_{h_1} \lesssim 5~\si{MeV}$.

Future experiments at the planning Higgs factories, such as CEPC~\cite{CEPCStudyGroup:2018ghi}, FCC-ee~\cite{FCC:2018byv}, and ILC~\cite{Baer:2013cma}, would be rather sensitive to $\kappa_Z$, $\kappa_W$, $\kappa_b$, and $\Gamma_{h_1}$.
For instance, the $1\sigma$ precision of the CEPC measurements on these quantities are estimated to be $\delta\kappa_Z = 0.25\%$, $\delta\kappa_W = 1.4\%$, $\delta\kappa_b = 1.3\%$, and $\delta\Gamma_{h_1} = 2.8\%$~\cite{CEPCStudyGroup:2018ghi}.
Figure~\ref{fig:Gammah1_kappaZ} also shows the expected 95\% C.L. coverage of the CEPC experiment on $\kappa_Z$ and $\Gamma_{h_1}$, and a large fraction of the selected parameter points would be properly tested.

Moreover, the CEPC project could also constrain the invisible Higgs decay branching ratio $\mathrm{BR}_\mathrm{inv}$ down to $0.3\%$ at 95\% C.L.~\cite{CEPCStudyGroup:2018ghi}.
In this model, we have nonzero $\mathrm{BR}_\mathrm{inv} = \Gamma(h_{1} \to \chi\chi)/\Gamma_{h_1}$ for $m_\chi < m_{h_1}/2$.
Figure~\ref{fig:mchi_BRinv} displays the selected parameter points projected onto the $m_\chi$-$\mathrm{BR}_\mathrm{inv}$ plane.
Current LHC data allow the parameter points with $\mathrm{BR}_\mathrm{inv} \lesssim 14\%$, while the CEPC experiment could probe most of the parameter points with $m_\chi < m_{h_1}/2$.

\begin{figure}[!t]
\centering
\subfigure[$\Omega_{\chi} h^2$-$\left< \sigma_\mathrm{ann} v \right>_\mathrm{FO}$ plane\label{fig:Omega_svFO}]
{\includegraphics[width=0.48\textwidth]{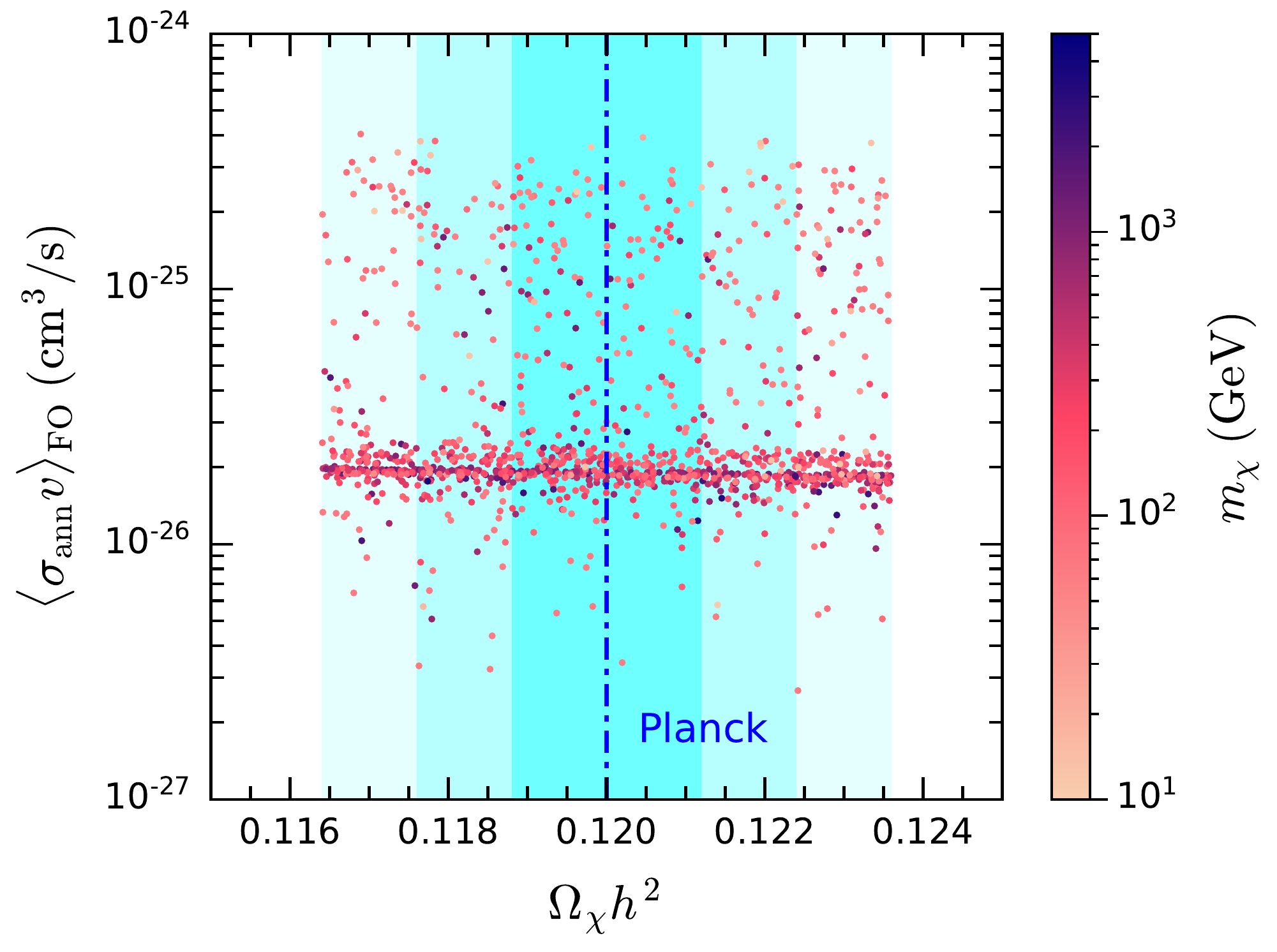}}
\hspace{.01\textwidth}
\subfigure[$m_\chi$-$\left< \sigma_\mathrm{ann} v \right>_\mathrm{D}$ plane\label{fig:mchi_svD}]
{\includegraphics[width=0.48\textwidth]{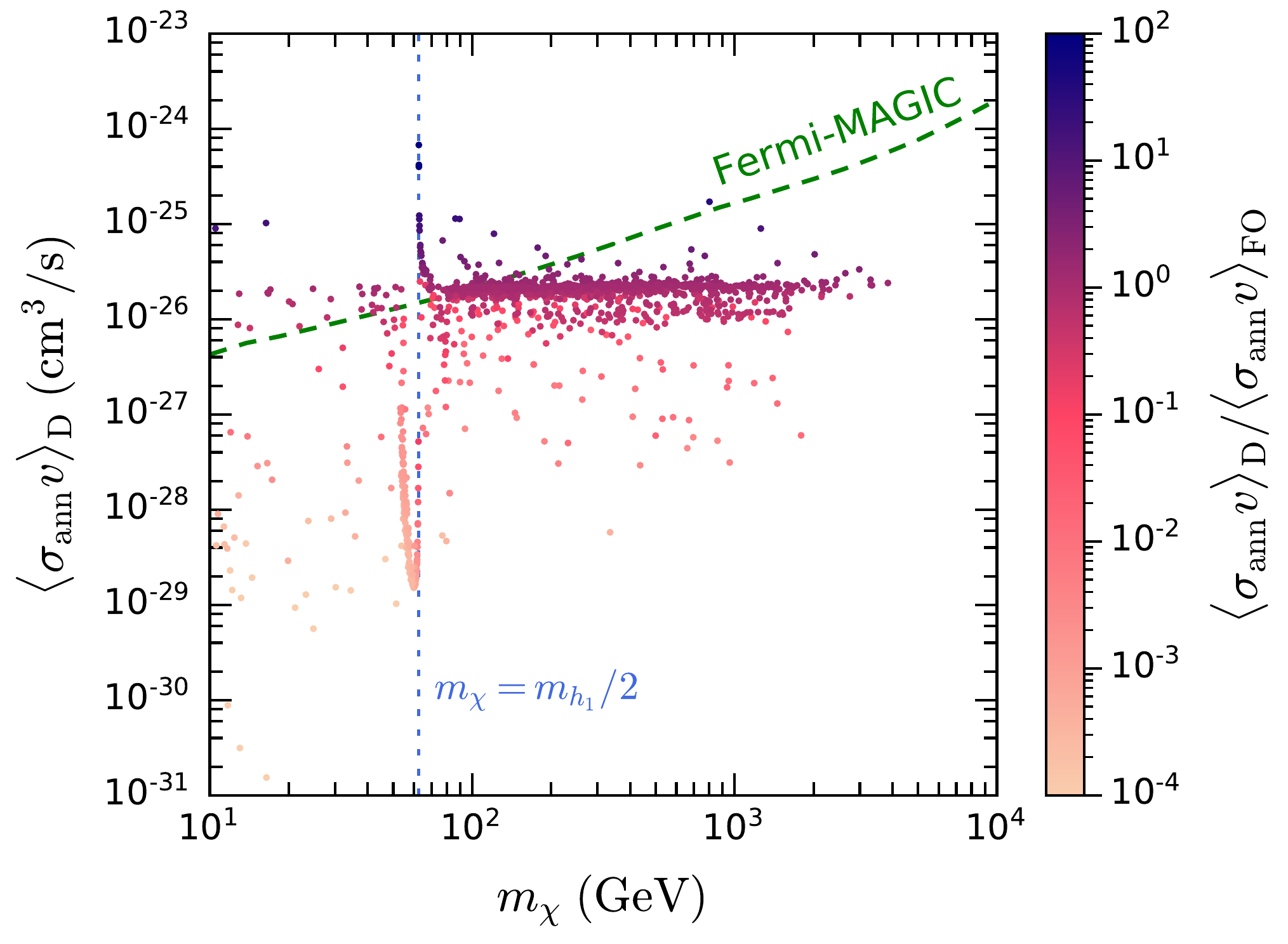}}
\caption{Selected parameter points projected onto the $\Omega_{\chi} h^2$-$\left< \sigma_\mathrm{ann} v \right>_\mathrm{FO}$ (a) and $m_\chi$-$\left< \sigma_\mathrm{ann} v \right>_\mathrm{D}$ (b) planes, with color axes corresponding to $m_\chi$ and $\left< \sigma_\mathrm{ann} v \right>_\mathrm{D}/\left< \sigma_\mathrm{ann} v \right>_\mathrm{FO}$, respectively.
The blue dot-dashed line and the colored regions in the left panel indicate the central value and the $1\sigma$, $2\sigma$, and $3\sigma$ ranges of the Planck measured relic abundance $\Omega_\mathrm{DM} h^2 = 0.1200\pm 0.0012$~\cite{Planck:2018vyg}.
In the right panel, the green dashed line denotes the upper limits from the Fermi-MAGIC $\gamma$-ray observations of dwarf galaxies at $95\%$ C.L.~\cite{MAGIC:2016xys}, while the blue dotted line corresponds to $m_\chi = m_{h_1}/2$.}
\end{figure}

In Fig.~\ref{fig:Omega_svFO}, the selected parameter points are presented in the $\Omega_{\chi} h^2$-$\left< \sigma_\mathrm{ann} v \right>_\mathrm{FO}$ plane, with a color axis indicating the pNGB WIMP mass $m_\chi$ and colored regions corresponding to the $1\sigma$, $2\sigma$, and $3\sigma$ ranges of the relic abundance $\Omega_\mathrm{DM} h^2 = 0.1200\pm 0.0012$ measured by the Planck experiment~\cite{Planck:2018vyg}.
The majority of the parameter points gather around the standard annihilation cross section $\left< \sigma_\mathrm{ann} v \right>_\mathrm{FO} \sim 2\times 10^{-26}~\si{cm^3/s}$.
The rest points with nonstandard freeze-out annihilation cross sections should arise from resonance or threshold effects of specific annihilation channels that leads to velocity-dependent cross sections~\cite{Griest:1990kh}.

The selected parameter points are further shown in the $m_\chi$-$\left< \sigma_\mathrm{ann} v \right>_\mathrm{D}$ plane in Fig.~\ref{fig:mchi_svD}, where the color axis denotes the ratio of $\left< \sigma_\mathrm{ann} v \right>_\mathrm{D}$ to $\left< \sigma_\mathrm{ann} v \right>_\mathrm{FO}$.
$\left< \sigma_\mathrm{ann} v \right>_\mathrm{D} \simeq \left< \sigma_\mathrm{ann} v \right>_\mathrm{FO}$ means that $\chi\chi$ annihilation is $s$-wave dominated, corresponding to the standard case.
The velocity dependence induced by the resonance or threshold effects would make $\left< \sigma_\mathrm{ann} v \right>_\mathrm{D}$ different from $\left< \sigma_\mathrm{ann} v \right>_\mathrm{FO}$.
It is obvious that the parameter points with $\left< \sigma_\mathrm{ann} v \right>_\mathrm{D} \neq \left< \sigma_\mathrm{ann} v \right>_\mathrm{FO}$ around the $m_\chi = m_{h_1}/2$ line in Fig.~\ref{fig:mchi_svD} are caused by the $h_1$ resonance effect, while the $h_2$ resonance effect leads to $\left< \sigma_\mathrm{ann} v \right>_\mathrm{D} \neq \left< \sigma_\mathrm{ann} v \right>_\mathrm{FO}$ for some of the rest parameter points.
The green dashed line in Fig.~\ref{fig:mchi_svD} indicates the $95\%$ C.L. upper limits of $\left< \sigma_\mathrm{ann} v \right>_\mathrm{D}$ from the Fermi-MAGIC observations of dwarf galaxies assuming a $b\bar{b}$ annihilation channel.
These limits can be approximately used to constrain the model.
We find that only a small fraction of the selected parameter points have been excluded.

\section{Conclusions and Discussions}
\label{sec:sum}

In this paper, we have constructed a UV-complete model for pNGB dark matter with a hidden $\UoneX$ gauge symmetry.
Two complex scalar fields $S$ and $\Phi$ carrying $\UoneX$ charges of 1 unit and 2 units are introduced.
The development of the $\Phi$ VEV $v_\Phi$ at a high scale breaks the $\UoneX$ gauge symmetry into an approximate $\UoneX$ global symmetry, which is softly broken by the $\mu_{S\Phi}$ term, leading to the desired pNGB WIMP DM setup.
As a result, the tree-level WIMP-nucleon scattering is suppressed by the UV scale $v_\Phi$.
We have found a scaling relation $\sigma_{\chi N}^\mathrm{SI} \propto v_\Phi^{-4}$, and hence $v_\Phi \gtrsim 10^{5}$ is high enough to escape direct detection.

Compared to the UV-completion with the $\Uonebml$ gauge symmetry~\cite{Abe:2020iph, Okada:2020zxo}, our model do not need to introduce right-handed neutrinos for anomaly cancellation.
Moreover, since the SM fermions do not carry $\UoneX$ charges, the interactions leading to WIMP decays are reduced.
Specifically, the interactions that induce WIMP decays only originate from the kinetic mixing between the $\UoneX$ and $\UoneY$ gauge fields.
This would relatively relieve the WIMP lifetime constraint on the UV scale $v_\Phi$.

A random scan in the parameter space has been carried out to obtain the parameter points satisfying the phenomenological constraints from the WIMP lifetime, the 125~GeV Higgs measurements, the observed DM relic abundance, and indirect detection of WIMP annihilation.
We have found that the WIMP lifetime bound from the Fermi-LAT $\gamma$-ray observations has set a lower limit on the UV scale, $v_\Phi \gtrsim 10^{10}~\si{GeV}$, which is indeed looser than $v_\Phi \gtrsim \mathcal{O}(10^{11}\text{--}10^{13})~\si{GeV}$ in the $\Uonebml$ case estimated in Refs.~\cite{Abe:2020iph, Okada:2020zxo}.
The parameter points satisfying current LHC Higgs measurements have $U_{11} \gtrsim 0.9$, $3.3~\si{MeV} \lesssim\Gamma_{h_1} \lesssim 5~\si{MeV}$, and $\mathrm{BR}_\mathrm{inv} \lesssim 14\%$.
A large fraction of these parameter points could be properly tested by future Higgs factories.

Additional constraints on this model come from direct searches for the $h_2$ boson at the $13~\si{TeV}$ LHC from decay channels such as $h_2 \to ZZ$~\cite{CMS:2018amk,ATLAS:2020tlo}, $h_2 \to W^+ W^-$~\cite{CMS:2019bnu,ATLAS:2022eap}, $h_2 \to t\bar{t}$~\cite{CMS:2019pzc}, and $h_2 \to h_1 h_1$~\cite{ATLAS:2021ifb}.
By reinterpreting these constraints, some parameter points remaining in our scan may have been excluded.
Nonetheless, the $h_2$ couplings to the $W$ and $Z$  bosons and to the top quark are highly suppressed by the mixing parameter $U_{12}$.
Thus, we expect most of the parameter points are still available.

\begin{acknowledgments}

This work is supported in part by the National Natural Science Foundation of China under Grants No.~11875327, No.~11905300, and No.~11805288, the Fundamental Research Funds for the Central Universities, and the Sun Yat-Sen University Science Foundation.

\end{acknowledgments}

\bibliographystyle{utphys}
\bibliography{ref}
\end{document}